\documentclass[a4paper,11pt]{article}
\pdfoutput=1 
\usepackage{jinstpub} 
\usepackage{hepparticles}
\usepackage{heppennames}
\usepackage{threeparttable}
\usepackage{subcaption}
\usepackage{tikz}
\usepackage{hyperref}
\usepackage{siunitx}
\graphicspath{{figures/}}
\usepackage{graphicx}
\graphicspath{{figures/}}
\usepackage{wrapfig}
\usepackage{float}
\usepackage{caption}
\usepackage{hyperref}
\usepackage{tabularx}
\usepackage{array}
\usepackage{amsmath}
\usepackage{amssymb}
\usepackage{color}
\usepackage{mathtools}
\usepackage{slashed}
\usepackage{pdfpages}
\usepackage{listings}
\usepackage{siunitx}
\usepackage{latexsym}
\usepackage{multirow}
\usepackage{lineno}

\title{\boldmath Qualification study of SiPMs  on a large scale for the CMVD Experiment}
\author[a,b,1]{Mamta Jangra,\note{Corresponding author.}}
\author[a,b]{Raj Bhupen,}
\author[b]{Gobinda Majumder,}
\author[b]{Kiran Gothe,}
\author[b]{Mandar Saraf,}
\author[b]{Nandkishor Parmar,}
\author[b]{B. Satyanarayana,}
\author[b]{R.R. Shinde,}
\author[b]{Shobha K. Rao,}
\author[b]{Suresh S Upadhya,}
\author[c]{Vivek M Datar,}
\author[d]{Douglas A. Glenzinski,}  
\author[d]{Alan Bross,}
\author[d]{Anna Pla-Dalmau,}
\author[e]{Vishnu V. Zutshi,}
\author[f]{Robert Craig Group,}
\author[f]{and E Craig Dukes,}

\affiliation[a]{Homi Bhabha National Institute, Mumbai-400094, India}
\affiliation[b]{Tata Institute of Fundamental Research, Mumbai-400005, India}
\affiliation[c]{The Institute of Mathematical Sciences, Chennai-600113, India}
\affiliation[d]{Fermi National Accelerator Laboratory, IL 60510, United States}
\affiliation[e]{Northern Illinois University, IL 60510, United States}
\affiliation[f]{Virginia University, VA, United States}
\emailAdd{mamta.jangra@tifr.res.in}

\abstract{A Cosmic Muon Veto (CMV) detector using extruded plastic scintillators is being designed around the mini-Iron Calorimeter (mini-ICAL) detector at the transit campus of the India based Neutrino Observatory, Madurai for the feasibility study of shallow depth underground experiments. The scintillation signals that are produced in the plastic due to muon trajectories are absorbed by wavelength shifting (WLS) fibres. The WLS fibres re-emit photons of longer wavelengths and propagate those to silicon photo-multipliers (SiPMs). The SiPMs detect these photons, producing electronic signals. The CMV detector will use more than 700 scintillators to cover the mini-ICAL detector and will require around 3000 SiPMs. The design goal for the cosmic muon veto efficiency of the CMV is $>$99.99\%. Hence, every SiPM used in the detector needs to be tested and characterised to satisfy the design goal of CMV. A mass testing system was developed for the measurement of gain and choice of the overvoltage ($V_{ov}$) of each SiPMs using an LED driver. The $V_{ov}$ is obtained by studying the noise rate, the gain of the SiPM. This paper describes the experimental setup used to test the SiPMs characteristics along with detailed studies of those characteristics as a function of temperature.}

\begin{document}
\maketitle
\flushbottom
\raggedbottom
\section{Introduction}
\label{intro}
A 51\,kton magnetized Iron Calorimeter (ICAL) was proposed~\cite{inoreport} at the underground laboratory of the India-based Neutrino Observatory (INO) to precisely measure the parameters of atmospheric neutrinos, mainly to study the effect of matter on neutrino oscillations~\cite{inowhitepaper}. The underground laboratory (with a rock cover of more than 1\,km in all directions) along with ICAL is planned to be located in Bodi West hills at Theni (\ang{9;57;50.1}\,N, \ang{77;16;21.8}\,E), India. ICAL will consist of three modules, where each of the modules will contain 150 layers of Resistive Plate Chambers (RPCs) interfaced between 5.6\,cm thick iron plates. Due to the low interaction cross-section of neutrino, there are only a handful of neutrino events in a day whereas there are a large number of cosmic-ray muons $\sim$ $3 \times 10^{8}$ /day incident on the surface about a km above the proposed ICAL detector~\cite{muonfluxcalculation}.
These cosmic muons act as a huge unwanted background interfering with the detection of muons arising from neutrino interactions. Placing a neutrino detector under a rock cover of 1km in all directions reduces the cosmic muon flux by a factor of $\sim$ $10^{6}$. At a depth of $\sim$ 100\,m or so, muons will be suppressed by a factor of $\sim$ $10^{2}$. To achieve a reduction factor $10^{6}$, an active CMV system with an efficiency of $\geq$ 99.99\% must be built around such a shallow depth detector to veto events arising from cosmic ray muons.
The prototype detector of the ICAL i.e., mini-ICAL is currently in operation at IICHEP, Madurai~\cite{gmsir}. The mini-ICAL consists of wenty $2\,m \times 2\,m$ glass RPCs sandwiched between 11 layers of 5.6 cm thick soft iron plates in the central $2\,m \times 4\,m$ region. The detector is magnetised using two sets of copper coils of 18 turns each. The central region of the mini-ICAL, where the RPCs are placed, has a nearly uniform magnetic field of 1.4\,T for a coil current of $\sim$ 900 amps. To suppress the cosmic muon background, it is planned is to cover the top and three sides of the mini-ICAL with an active veto detector for the detection of the cosmic ray muons and estimate the CMV efficiency of the detector, using the same concept as the Mu2e collaboration~\cite{mu2etdr}. The goal is to build an active veto detector with an efficiency of 99.99\% and a false-positive rate of less than $10^{-5}$.

The CMV detector for mini-ICAL will have four layers of extruded scintillators~\cite{lowcost} of size 450\,cm $\times$ 5\,cm $\times$ 2\,cm on top and three layers of 460\,cm $\times$ 5\,cm $\times$ 1\,cm on three sides of the mini-ICAL~\cite{mamta1}. The fourth side is planned to not be covered with veto layers due to the maintenance and troubleshooting of the mini-ICAL. In all, the veto system will comprise more than 700 extruded plastic scintillators. The surface area of each layer on top is ($450\,cm \times 440\,cm$) and is ($460\,cm \times 200\,cm$) for each of the side layers. The scintillation signal will be collected through WLS fibres ~\cite{kuraray} embedded in the extruded scintillator and the light from both ends of the fibres will be readout using SiPMs.

This paper describes the test setup to characterise $\sim$ 3500 SiPMs. These studies were performed as a part of the R$\&$D program for the CMV system around mini-ICAL. Along with the basic performance studies we have also looked in detail at the effect of temperature on the SiPM and whether a feedback circuit is required for $V_{bias}$ to compensate for the gain of SiPM due to the variation of temperature or not.

\section{Hamamatsu SiPM 13360-2050VE}
\label{hamamatsu_sipm}
\begin{figure} [htbp]
\centering
\includegraphics[width=0.5\textwidth]{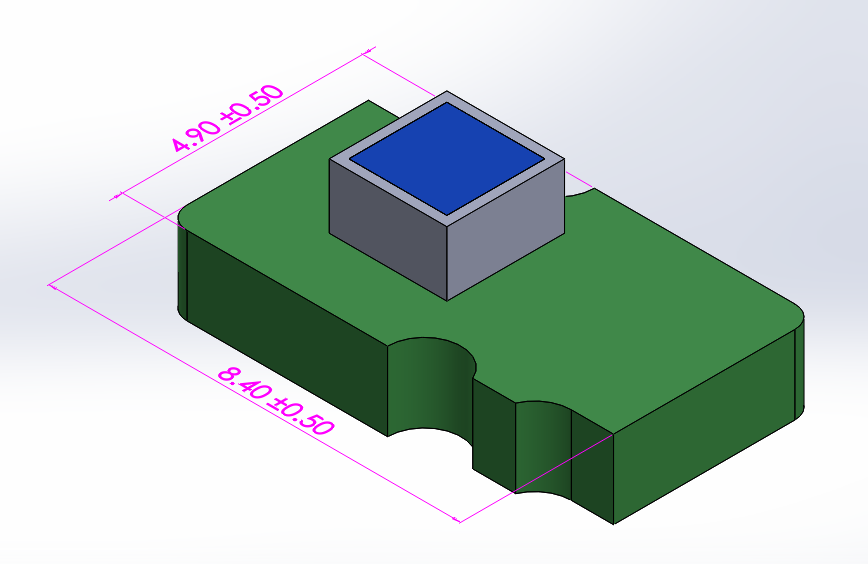}
\hspace{1cm}
\includegraphics[width=0.35\textwidth]{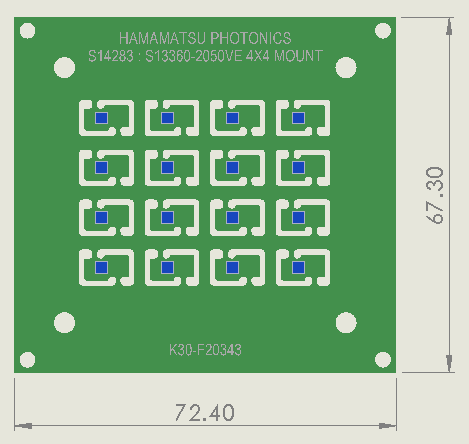}
\caption{Geometrical structure of 16 SiPMs in a panel on right, sim like structure of a SiPM on left.}
\label{fig:sipmpack}
\end{figure}

The SiPM selected for the CMVD experiment is the Hamamatsu S13360-2050VE. This SiPM model has an effective photosensitive area of  2\,mm$\times$2\,mm and a total of 1584 square microcells with a microcell pitch of 50\,$\mu$m, fill factor of 74\%, the breakdown voltage of (53 $\pm$ 5)\,V at room temperature~\cite{sipmspecs}. The 2\,mm$\times$2\,mm active area of the chosen SiPM (S13360-2050VE) makes it very suitable for coupling with the WLS fibre of 1.4\,mm diameter. The SiPMs are available mounted on tiny Carrier Boards (sim-like structure on left in figure~\ref{fig:sipmpack}) and 16 such carrier boards are present in a panel as shown on the right in figure~\ref{fig:sipmpack}.

\section{Test Setup for mass testing of the SiPMs}
\label{test_setup}
The major components used for SiPM testing are LED driver as a light source, DRS for front end readout, and Transimpedance (TI) amplifiers to amplify the raw SiPM signals. Before going to the overall setup, here are some of the details about these major components.

\subsection{Pulsed LED for characterising the SiPMs}
\label{pulsed_led}
The standalone SiPMs were tested and characterised using an SP5601 CAEN LED system~\cite{CAEN}. The LED system consists of an ultrafast LED driver with tunable periodicity and intensity and a pulse width of a few ns. It can provide a light burst with a tunable intensity varying from a few to more than tens of photons required for calibration purposes.
\subsection{DRS4}
\label{drs_section}
The Domino Ring Sampler (DRS4) ASIC is a switched capacitor array (SCA) capable of sampling 9 differential input channels at a sampling speed of up to 5\,GS/s~\cite{DRS1}. The SCA consists of 1024 capacitive cells per channel to store the input analog waveform. The stored waveform samples can be readout via a shift register clocked at 33\,MHz for external digitization. For SIPM data collection we used the DRS4 evaluation board~\cite{DRS2}. The evaluation board has four analog inputs, a USB connection for data readout and board configuration and, on-board trigger logic. It is equivalent to a four channel 5\,GS/s digital oscilloscope. The Evaluation boards come with PC software to display the waveforms captured from the four-input channels as well as store the captured data in the local disc.
\subsection{Transimpedance Amplifier}
\label{trans_amplifier}
\begin{figure} [h!]  
\centering
\includegraphics[width=\textwidth]{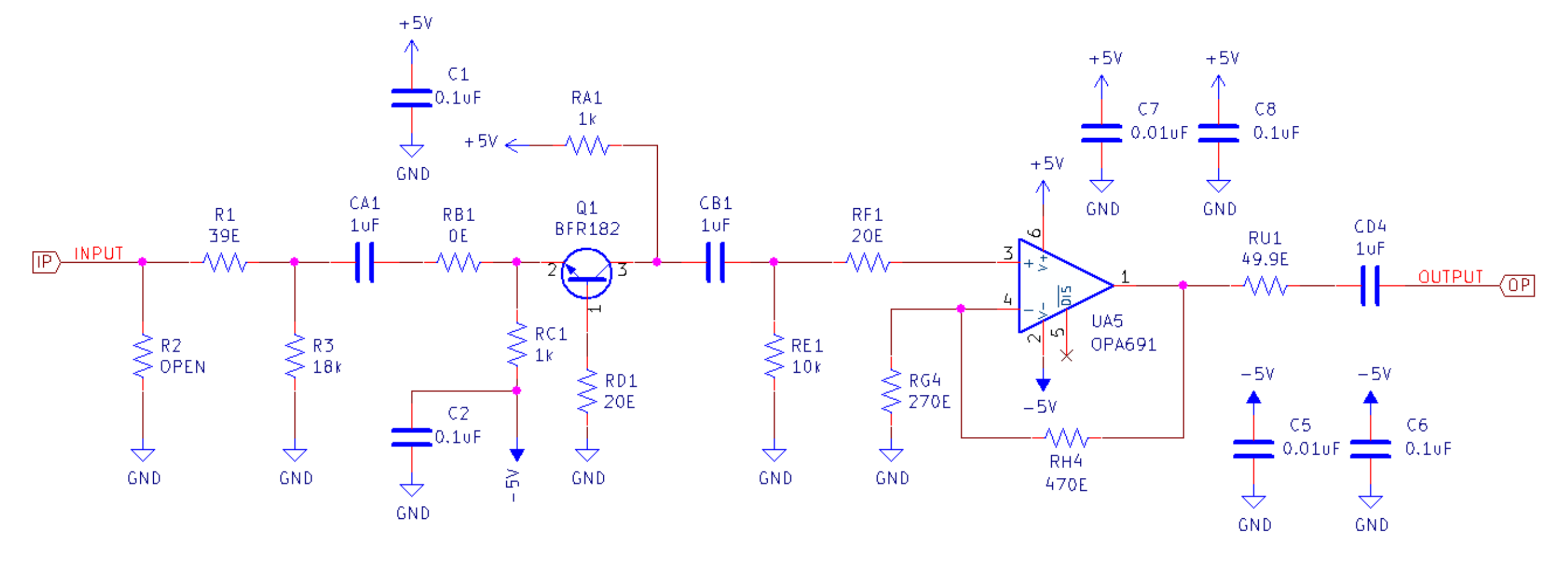}
\caption{Circuit for the Transimpedance Amplifier.}
\label{fig:Tamplifier}
\end{figure}

The SIPM generates a charge of around 0.27\,pC for every photoelectron (p.e.) at $V_{ov}$ of 2.5\,V. The current that flows due to this charge into a 50\,$\Omega$ termination resistor of an oscilloscope, generates a very small voltage pulse of the order of a few millivolts. To get reliable charge distribution data, a Trans-Impedance (TI) amplifier was used. The gain of the TI amplifier is determined by the collector resistor, RA1 in parallel combination with RE1 shown in figure~\ref{fig:Tamplifier}. The OpAmp stage
provides a further gain of 1.37 and helps in driving the coaxial cable.

\subsection{Overall Setup}
\label{overall_setup}
\begin{figure} [h!]  
\centering
\includegraphics[width=0.65\textwidth]{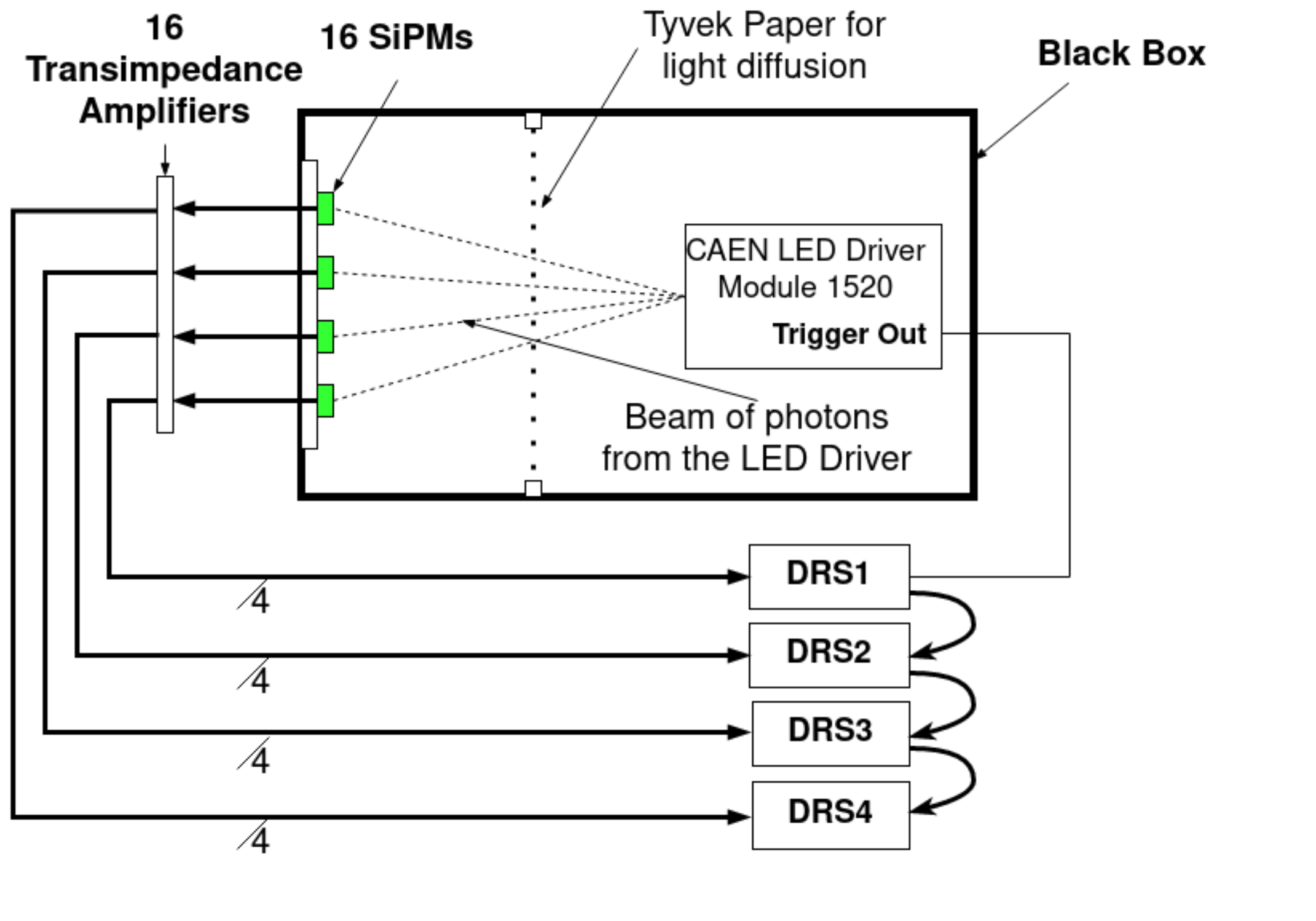}
\caption{Signal readout paths through the DRS system.}
\label{fig:Overallsetup}
\end{figure}

A wooden black box of size 120\,cm$\times$60\,cm was used for mass testing of the SiPMs with the LED source. The overall setup is shown in figure~\ref{fig:Overallsetup}. The test setup contains a motherboard to mount the 16 SiPM's panel (figure~\ref{fig:sipmpack}) which was fixed inside the black box. The SiPMs on the panel will have secure electrical connections using pogo (spring-loaded) pins on the motherboard. The LED driver is kept inside the black box such that the SiPM will directly face the LED flash. The LED pulse is synchronized with a trigger by an external pulser. A Tyvek paper was used as a screen between the LED driver and the SiPMs to diffuse the light. The intensity of the LED source and the distance from the SiPMs is adjusted so that a handful of photons are incident on every SiPM and the charge distribution can be seen clearly without saturating the system.\\

\begin{figure} [htbp]  
\centering  
\includegraphics[width=0.35\textwidth]{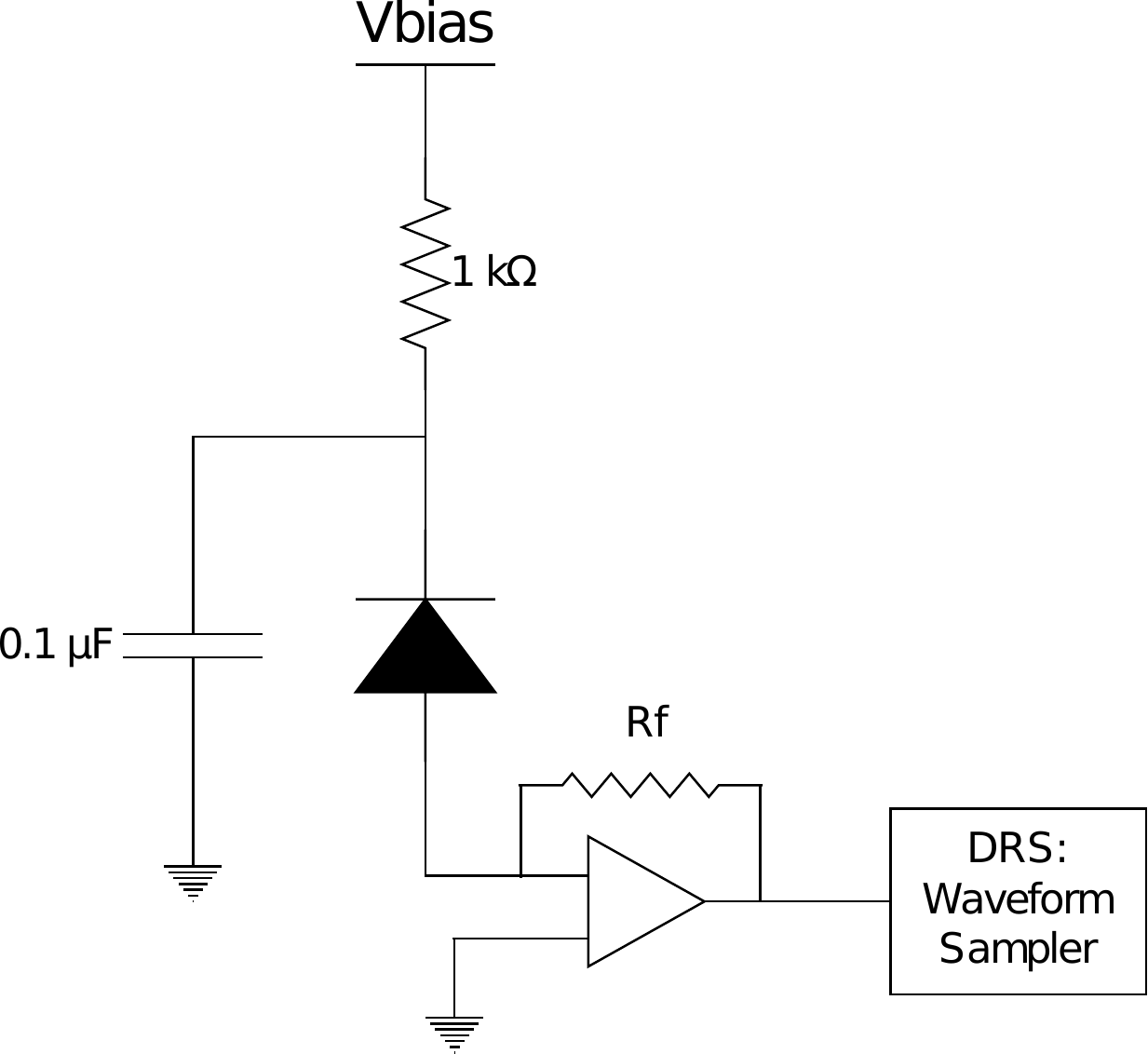}
\caption{Schematic diagram of SiPM test setup.}
\label{fig:sipmckt}
\end{figure}

When photons from the LED are incident on the SiPMs, electron-hole pairs will be generated in the corresponding microcells and an avalanche will be produced in every corresponding microcell due to the applied bias voltage ($V_{bias}$) leading to a current pulse in the circuit. For analog SiPMs (common readout for all microcells), the overall signal of the SiPM will be proportional to the number of individual microcell signals which are read. The SiPM signals are taken from the motherboard to the TI amplifiers using coaxial cables, the voltage signals (amplified SiPM signals) of the TI amplifiers are then connected to the DRS waveform sampler modules as it is shown in figure~\ref{fig:sipmckt}. The LED driver generates a trigger pulse when it fires the LEDs, this signal was used as a trigger input to the DRS modules, 4 of which were used to capture the signals of the 16 SiPMs. The charge generated by every SiPM was measured from the DRS data by using the equation~\ref{eqn:charge}

\begin{eqnarray}
q = \frac{1}{R\times G} \int_{to}^{t1} V(t) dt
\label{eqn:charge}
\end{eqnarray}            
where R is the input resistance of the transimpedance amplifier and $G$ is the gain of the operational
amplifier stage. 

\section{Data Analysis}
\label{data_analysis}
\begin{figure} [h!]  
\centering
\includegraphics[width=.9\textwidth]{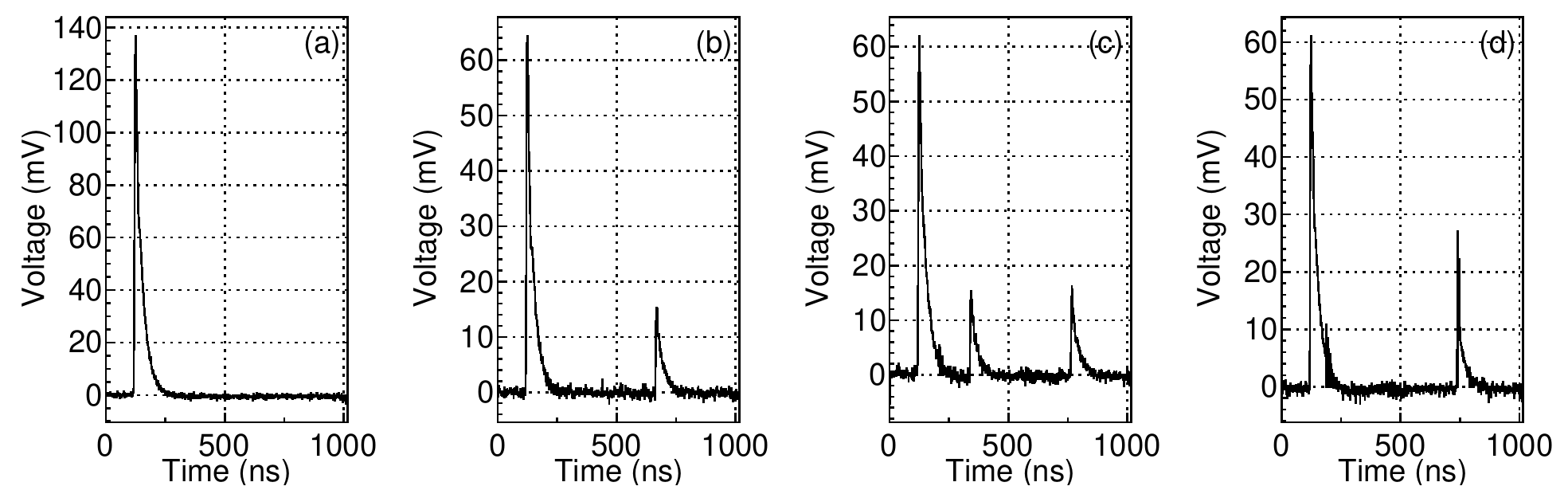}
\caption{Raw signals of SiPM; (a) Ideal clear signal of LED, (b) signal with one dark noise, (c) signal with two dark noises and (d) signal with optical crosstalk signal.}
\label{fig:signal}
\end{figure}

As discussed in the section\,\ref{test_setup}, there are 16 SiPMs carrier boards in one panel, and all those 16 SiPM are illuminated with LED at the same time and each of the SiPM signals
is amplified by a TI amplifier of gain 0.909\,mV/$\mu$A and the output voltage is stored through the DRS system. The sampling rate is 2\,GHz. Typical signals
of SiPM are shown in figure~\ref{fig:signal}. The first signal figure~\ref{fig:signal}\color{blue}(a)\color{black}\hspace{0.1cm} is the clean output correlated with the LED pulse. 
The rise time of the pulse is $\sim$6\,ns and the signal falls with a decay constant of $\sim$28\,ns. The rise time of SiPM pulse without an amplifier is $\sim$2\,ns, increase of rising time to 6\,ns is due to the effect of the TI amplifier, \color{black}whereas the decay constant mainly depends on the quenching resistance and associated capacitances. Signals in figure~\ref{fig:signal}\color{blue}(b)\color{black}\hspace{0.1cm} and~\ref{fig:signal}\color{blue}(c)\color{black}\hspace{0.1cm} shows the LED signal associated with single and double p.e. peaks due to single dark noise  and two dark noise events respectively. The LED signal may also be associated with signals from optical cross talk, as shown in figure~\ref{fig:signal}\color{blue}(d)\color{black}. The dark noise and optical crosstalk depend on the temperature and $V_{ov}$ \cite{mamta1}.

\begin{figure} [htbp]
\centering
\includegraphics[width=9.cm]{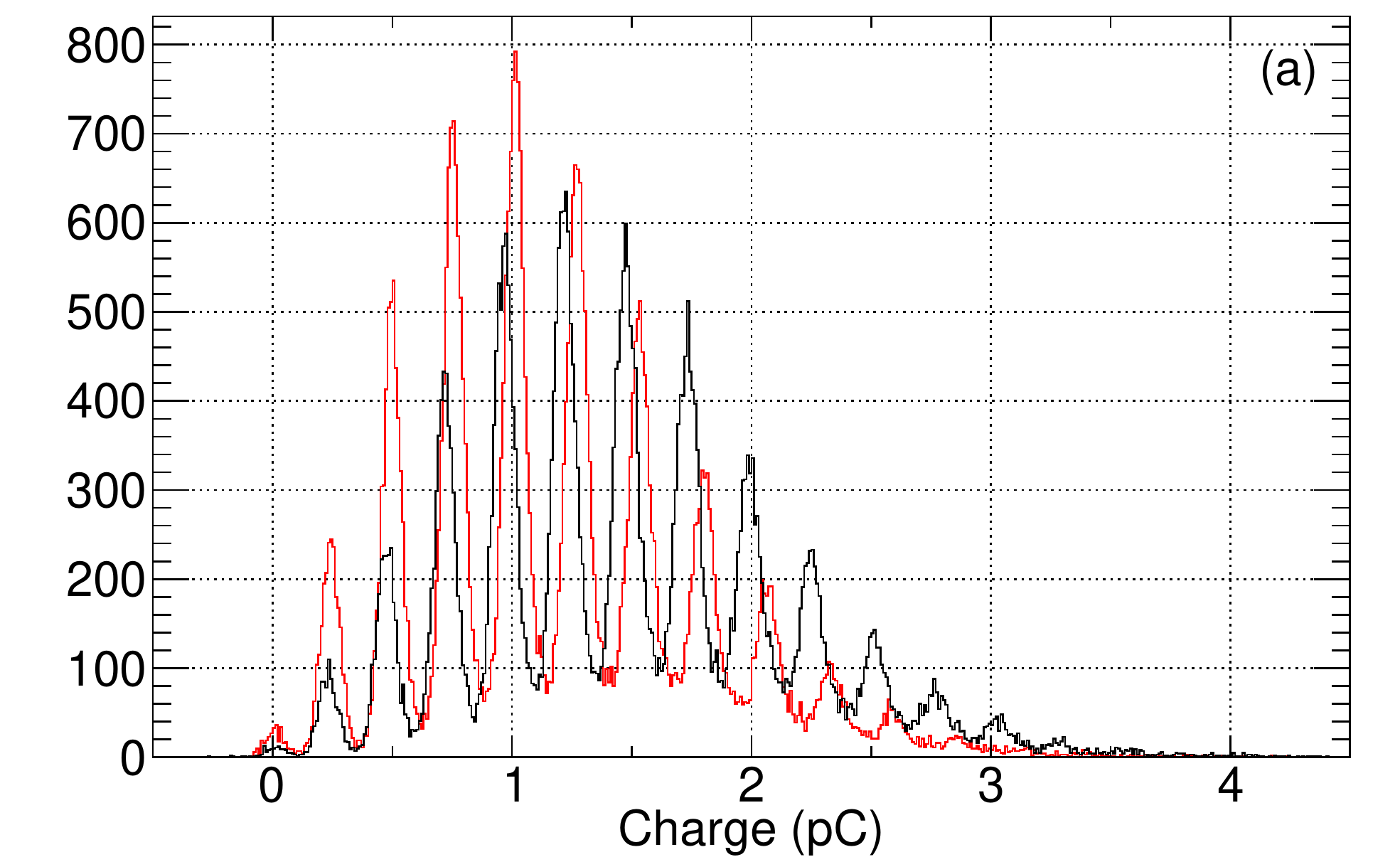}
\includegraphics[width=3.7cm]{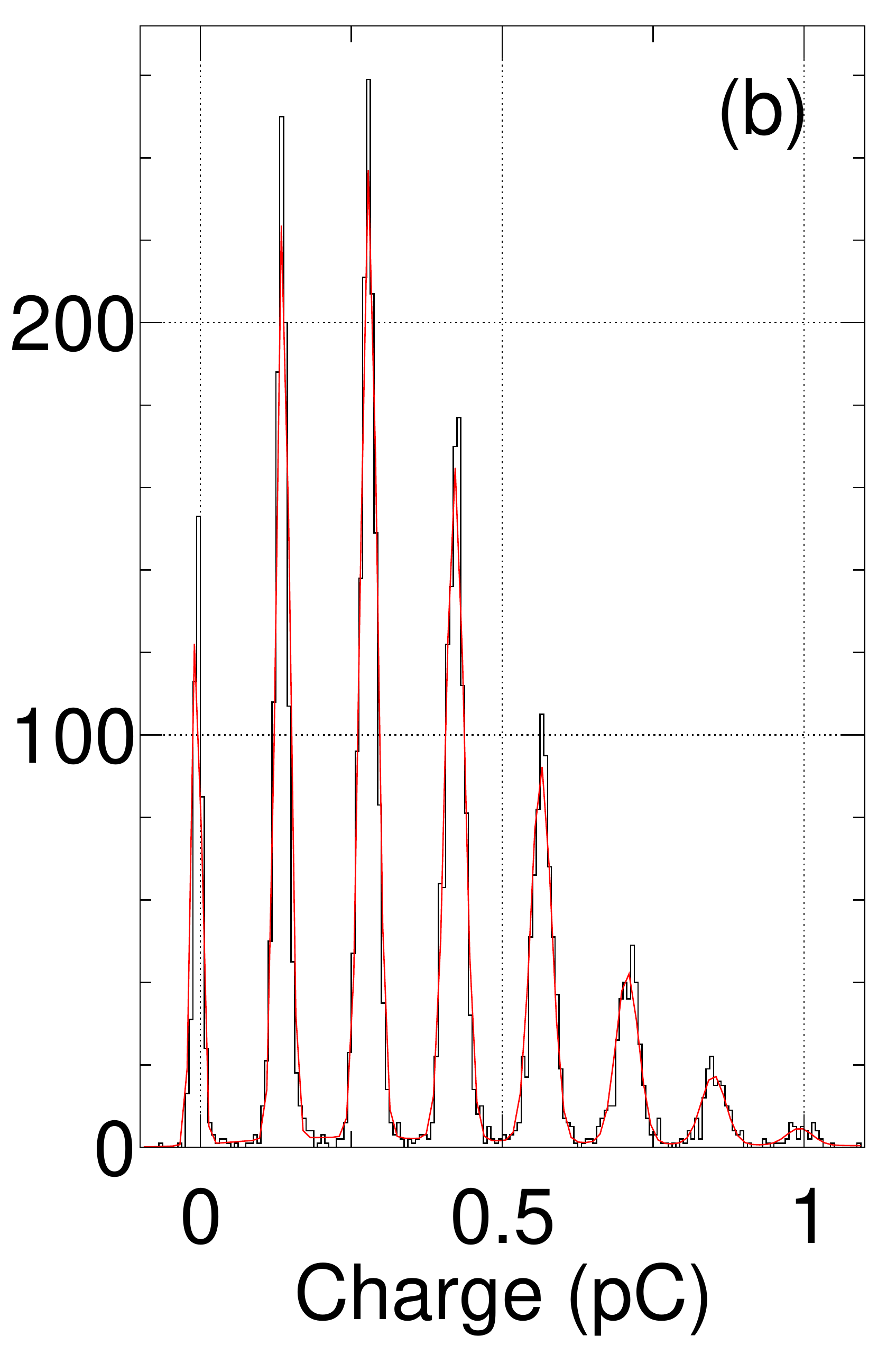}
\caption{Pedestal subtracted integrated signals due to LED pulse at at 54\,V for
        (a) two SiPM in a panel and (b) Fit of a spectrum to obtain gain.}
\label{fig:intg_chrg}
\end{figure}

A SiPM is characterised by its gain as a function of $V_{ov}$, its dependence on temperature and noise rate as a function of $V_{ov}$ and temperature. The gain of a SiPM is defined as the
integrated charge due to a single p.e. as defined in equation \ref{eqn:charge}. The integration time is taken as 100\,ns, which contains more than 95\% of the signal. This optimum width is
chosen to reduce the dark noise as well as accept most of the signal. 
The signals in figure~\ref{fig:signal} show the baseline of the electronic signal to be close to zero, but to avoid any baseline fluctuations, integrated pedestal values of the same 100\,ns width is subtracted from the signal. Typical integrated signals due to LED exposure are shown in figure~\ref{fig:intg_chrg}\color{blue}(a)\color{black}\hspace{0.1cm} for two SiPMs on a single panel. From this figure, it is clear that the gain of different SiPMs is not exactly the same, although the specifications are the same. The gain is quantified by fitting those distributions with the combination of Gaussian functions for each p.e. peak and a wider background function. A typical fit is shown in figure~\ref{fig:intg_chrg}\color{blue}(b)\color{black}. Details of the fit function are given in~\cite{mamta1}. As different peaks are due to different p.e., the gap between two peaks is the estimation of the gain of the SiPM at that $V_{ov}$ and the temperature.

\begin{figure} [h!]  
\centering
\includegraphics[width=0.9\textwidth]{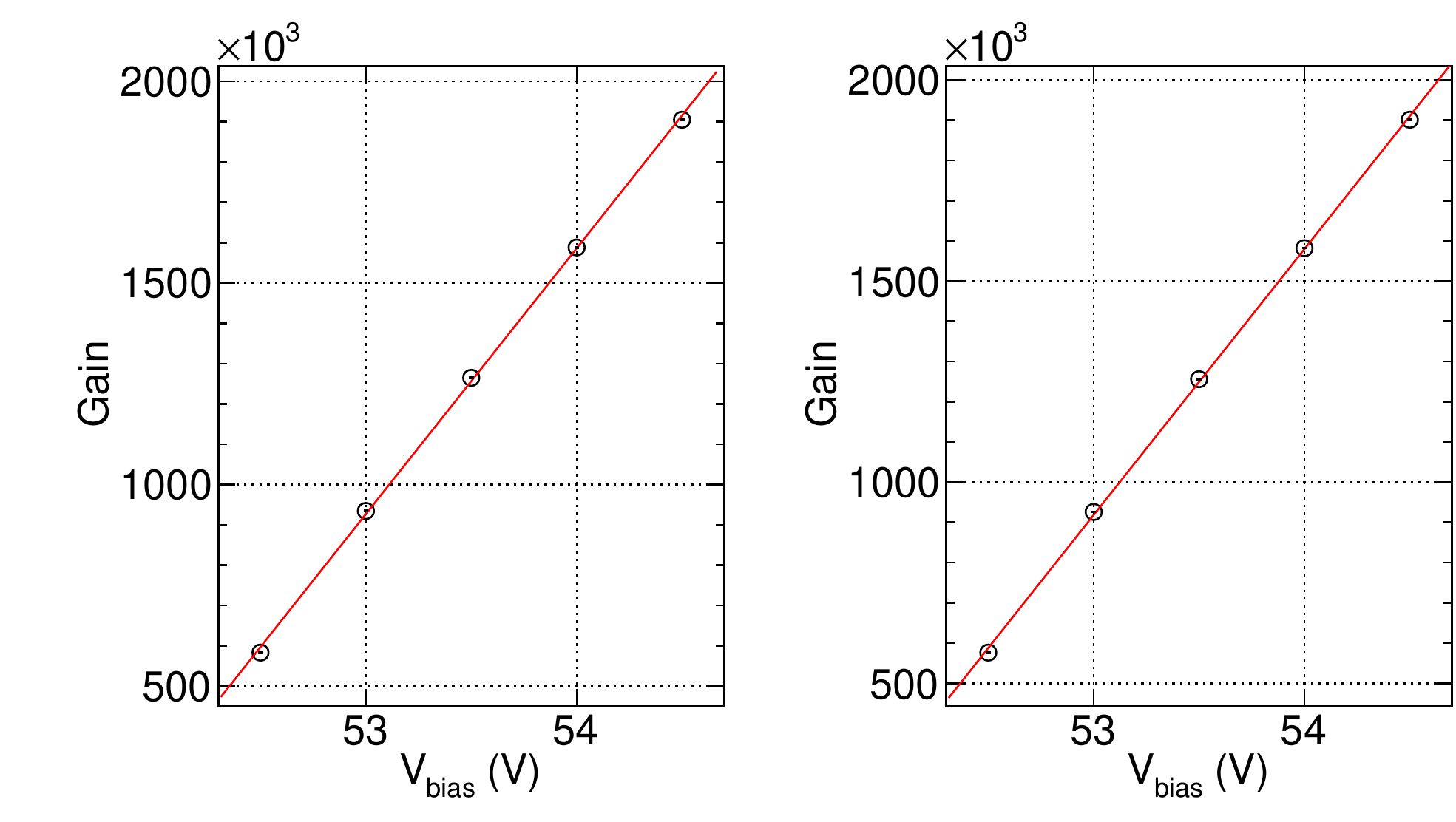}
\caption{Variation of gain as a function of applied HV along with the linear fit function for few SiPMs.}
\label{fig:gainhv}
\end{figure}

To obtain the threshold voltage ($V_{th}$) and the variation of gain with respect to applied HV
(dG/dV) data are collected for five different HV above the rated $V_{th}$. A typical distribution
of gain vs HV is shown in figure~\ref{fig:gainhv}. The points are fitted with a simple linear function,
where the intercept of the line in the voltage axis is defined as $V_{th}$ and the slope, (dG/dV) is the measure of the variation of gain with the change in overvoltage, $V_{ov}$, the difference between the $V_{bias}$ and  $V_{th}$. During this data taking, the temperature was not controlled properly, but the room temperature was 24$\pm$1$^{\circ}$\,C. The effect of
temperature will be discussed in detail in section \ref{temp_depend}.

\begin{figure} [h!]  
\centering
\includegraphics[width=0.8\textwidth]{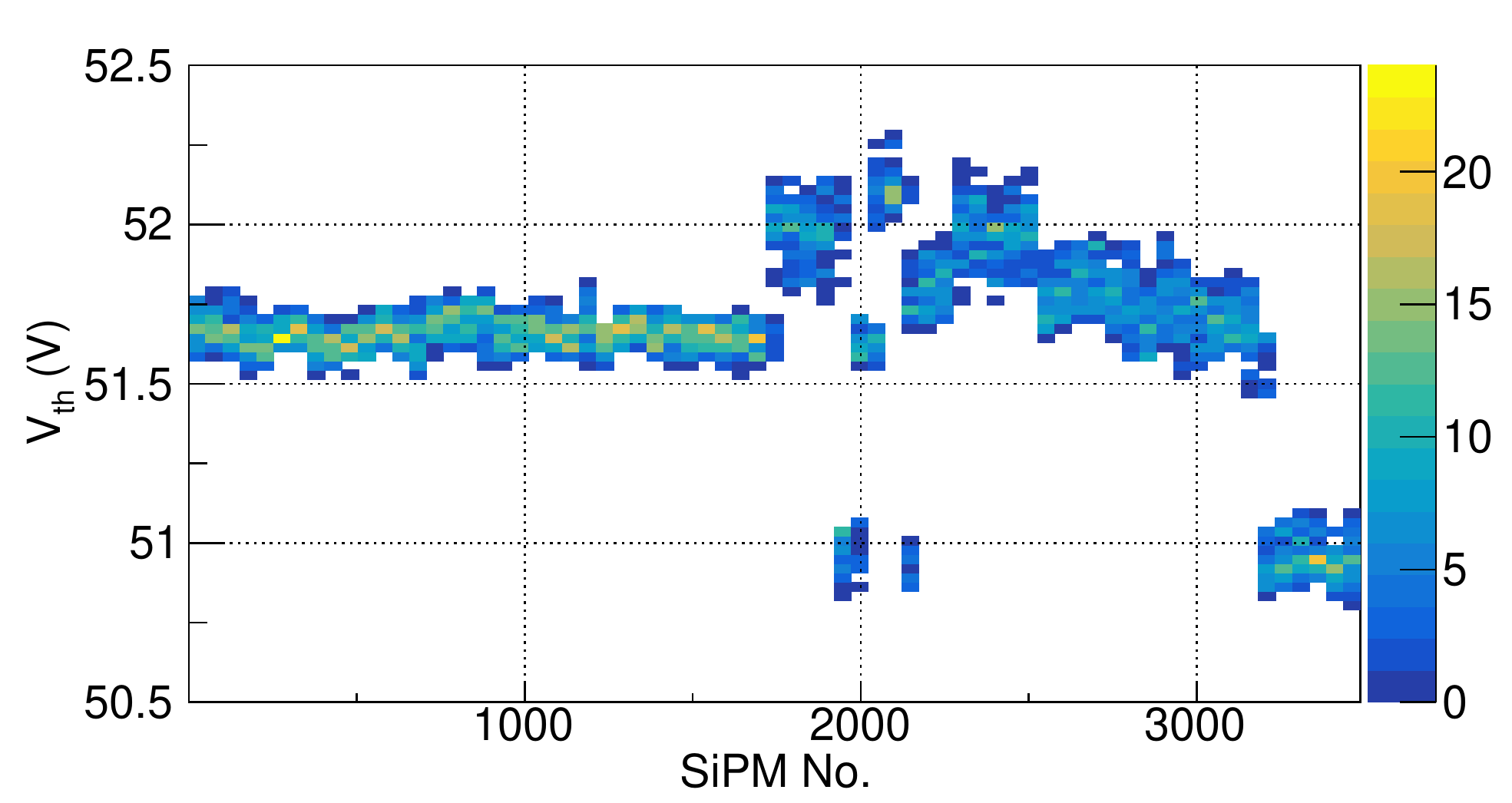}
\caption{$V_{th}$ of all tested SiPMs for CMVD.}
\label{fig:vbdnumb}
\end{figure}

There were a total of 220 panels with 16 SiPMs each. The SiPMs of one of the panels were used before the testing, thus we have tested all the SiPMs in the remaining 219 panels. The $V_{th}$ of all SiPMs are shown in figure~\ref{fig:vbdnumb}.
The $V_{th}$ varies from 50.7\,V to 52.3\,V and matches with the specification given by
Hamamatsu, at 25\,$^\circ$\,C. The point to mention is that the $V_{th}$ of SiPMs in each panel are
nearly the same and the maximum deviation is 0.5\,V.

\begin{figure} [h!]  
\centering
\includegraphics[width=0.8\textwidth]{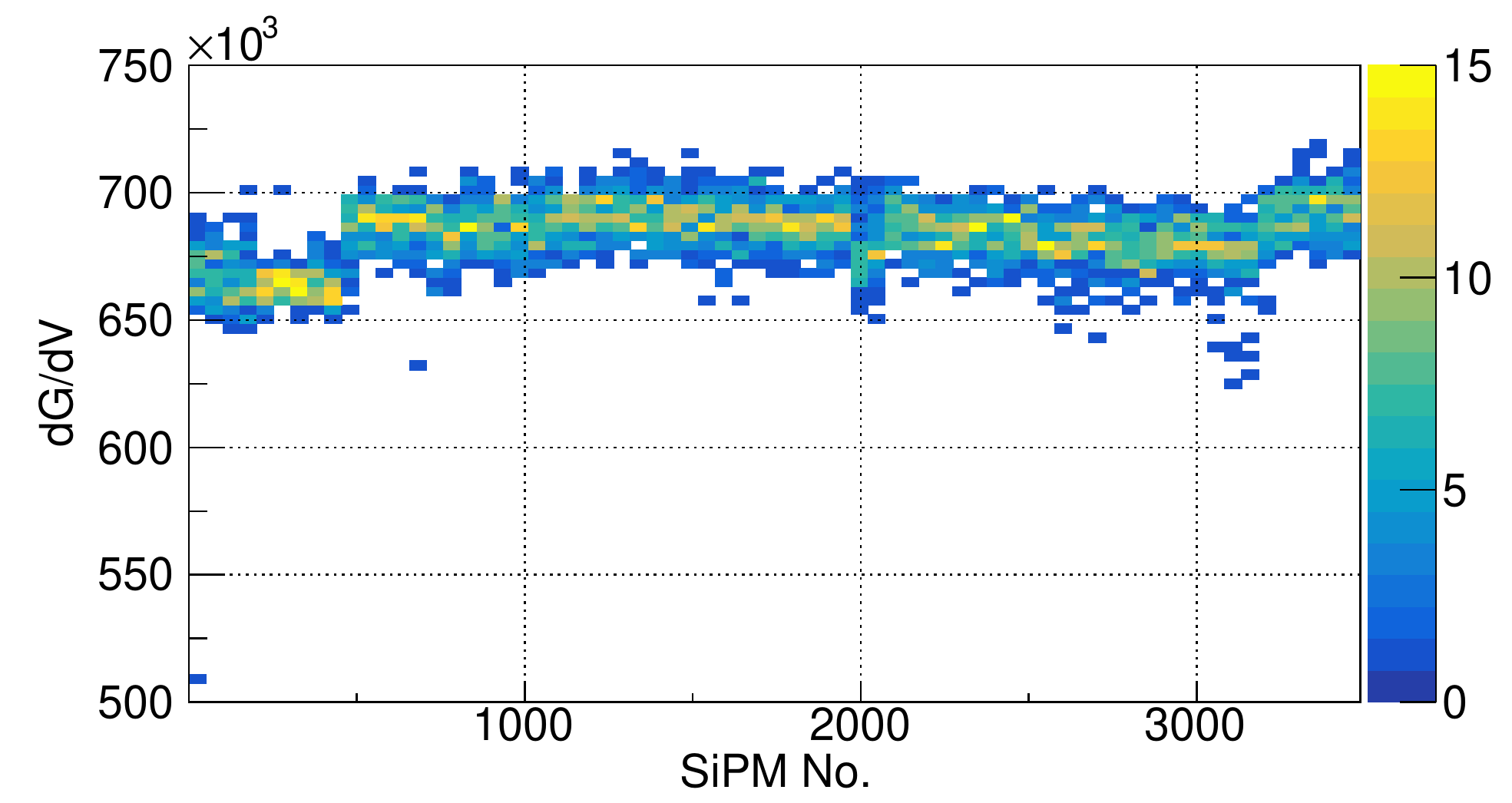}
\caption{dG/dV of all tested SiPMs for CMVD.}
\label{fig:dgdvnumb}
\end{figure}

The variation of gain with respect to applied voltage ($dG/dV$) about the $V_{th}$ is shown
in figure \ref{fig:dgdvnumb}. Most of the SiPMs are having $dG/dV \sim 6.8\times10^{5}$. There
are few outliers, but only one SiPM has much lower gain, $5.1\times10^{5}$.

\begin{figure} [h!]  
\centering
\includegraphics[width=.8\textwidth]{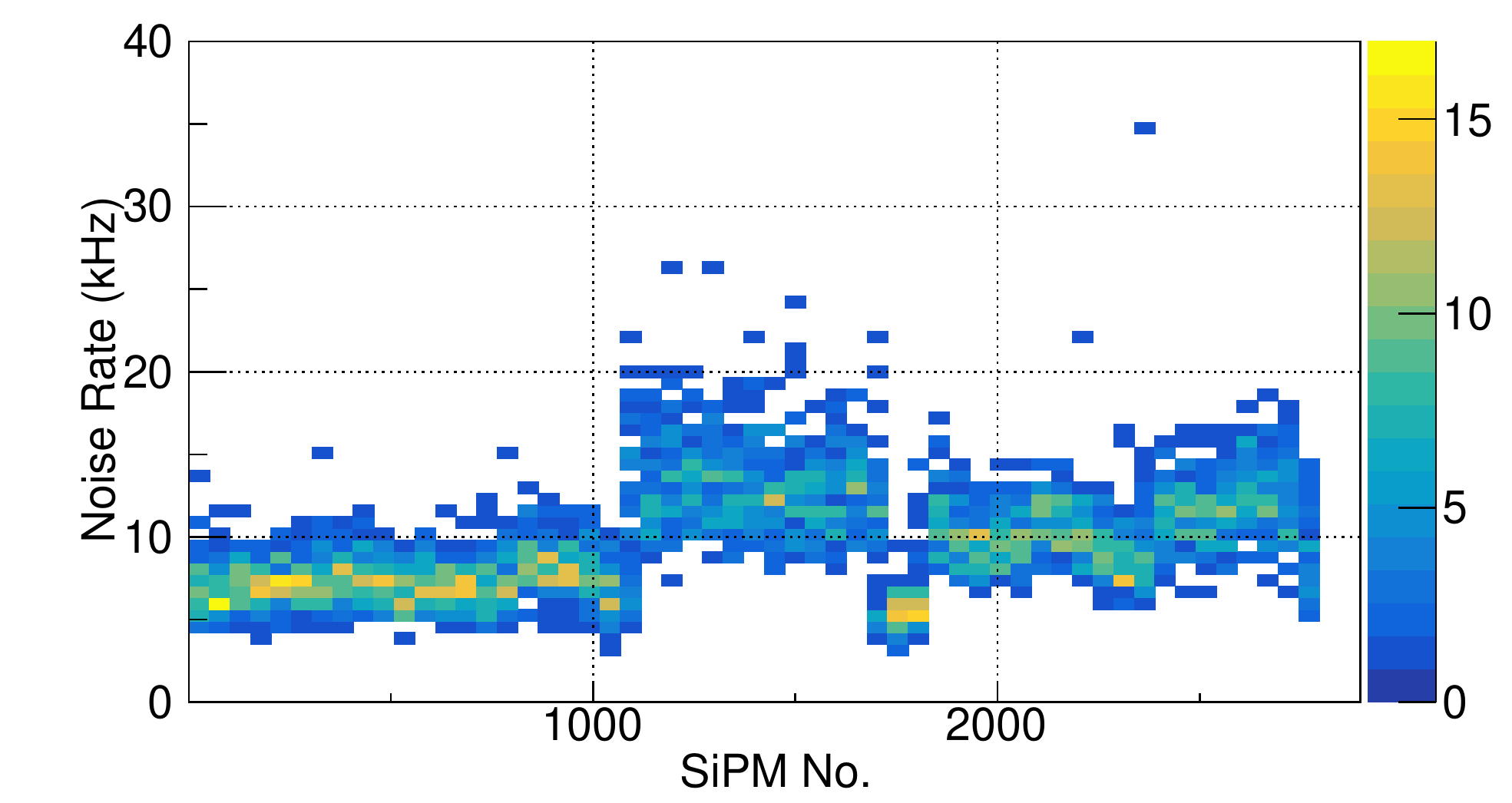}
\caption{Noise rate of all tested SiPM for $V_{ov}$ 3\,V and $Q_{th}$ 0.48\,pC.}
\label{fig:noiserate}
\end{figure}

From the figures~\ref{fig:vbdnumb} and ~\ref{fig:dgdvnumb} it is clear that all except one SiPM (low gain with $dG/dV \sim 5.1\times10^{5}$) is suitable for the CMV, where gains in all SiPMs are reasonably high and the variation of $V_{th}$ is moderate. The third crucial parameter of the SiPM is dark noise. That was estimated by integrating the SiPM signal by randomly triggered 100\,ns time windows. All these data were taken without
any LED source and also at $V_{th}\sim$ 3\,V. The random noise is calculated by counting the fraction of events with an integrated signal more than 0.48\,pC, which is equivalent to $\sim$ 1.5\,p.e.. 
 The variation of noise rate for all these SiPM is shown in figure~\ref{fig:noiserate}. The maximum noise rate at 1.5\,p.e. threshold is 34.9\,kHz for $V_{th}$=3\,V, which is well below the acceptance
threshold~\cite{mamta1}.

\begin{figure} [htbp]  
\centering
\includegraphics[width=0.95\textwidth]{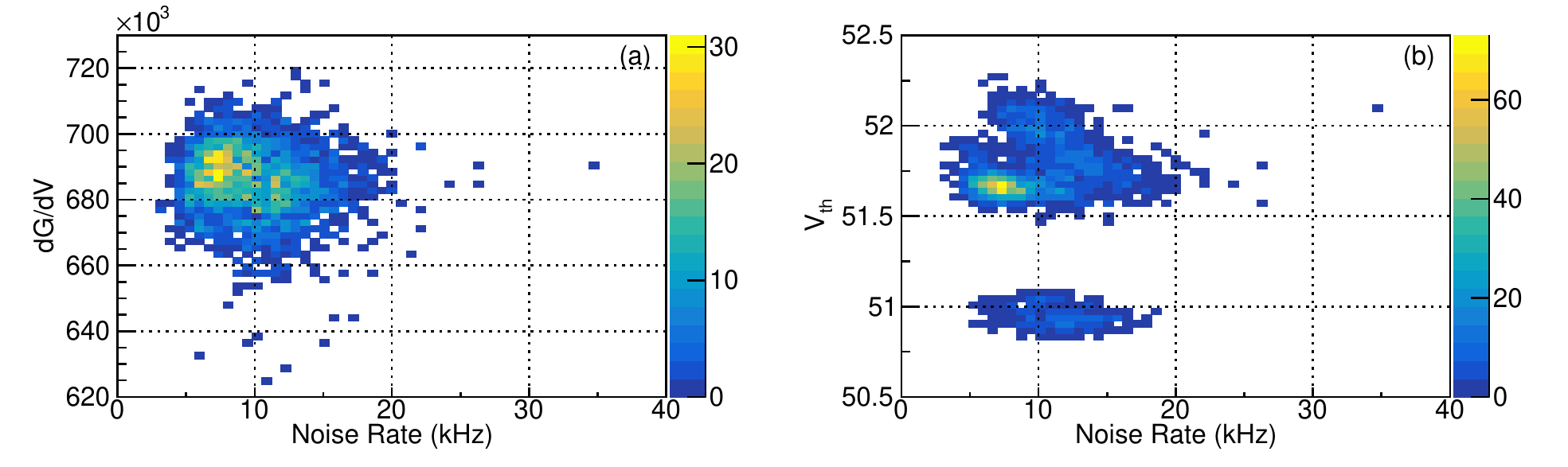}
\caption{Correlation between the noise rate and a) $dG/dV$, b) $V_{th}$.}
\label{fig:dgdv_vth_noise}
\end{figure}

Based on the criteria and the noise rate of individual SiPM, all SiPM are satisfying the requirements of the CMVD. The figure \ref{fig:dgdv_vth_noise} shows the correlation of noise rate with $dG/dV$ on left and $V_{th}$ on right respectively. There is no observed
correlation of noise with those two parameters.

\section{Setup for Controlled Temperature Testing}
\label{controlled_temp}
\begin{figure} [htbp]  
\centering
\includegraphics[width=0.7\textwidth]{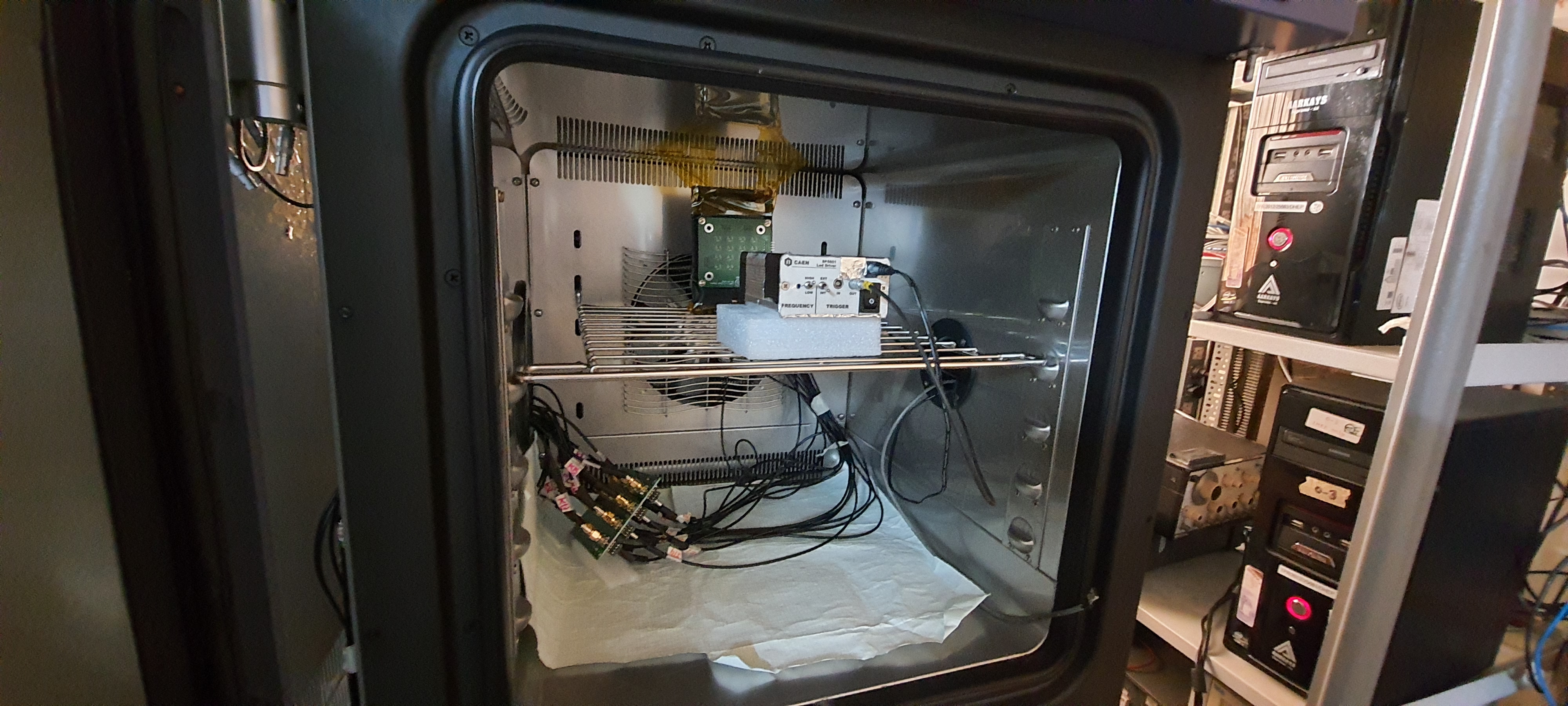}
\caption{Setup inside controlled temperature chamber.}
\label{fig:tempset}
\end{figure}

SiPMs show a large variation of their gain on the operational temperature. As per the SiPM datasheet, the temperature coefficient of the operating voltage is 54\,mV/$^\circ$\,C. To measure this value, it was necessary to take the setup shown in the figure~\ref{fig:Overallsetup} above to a chamber where the temperature could be maintained at a known value. We used the thermal chamber, ``SU-642'' from ``ESPEC'' for this purpose. A view of our set up inside the chamber is shown in figure~\ref{fig:tempset}. This chamber can maintain a temperature from $-$20 to 80$^\circ$\,C with a temperature fluctuation of $\pm$0.3$^{\circ}$\,C~\cite{tempspec}, but during the data taking period, the temperature is varied from $-$20 to 50$^\circ$\,C.

\section{Variation of the gain of SiPM with temperature}
\label{temp_depend}
For this study, we have used only 4 SiPMs of one panel. The temperature of the chamber is varied from $-20^\circ$\,C to +50$^\circ$\,C and for each temperature, LED data are taken for around 20 voltages. 
  There are many data points around 25$^\circ$\,C, which is the average operating temperature in the IICHEP laboratory.

\begin{figure} [htbp]  
\centering
\includegraphics[width=\textwidth]{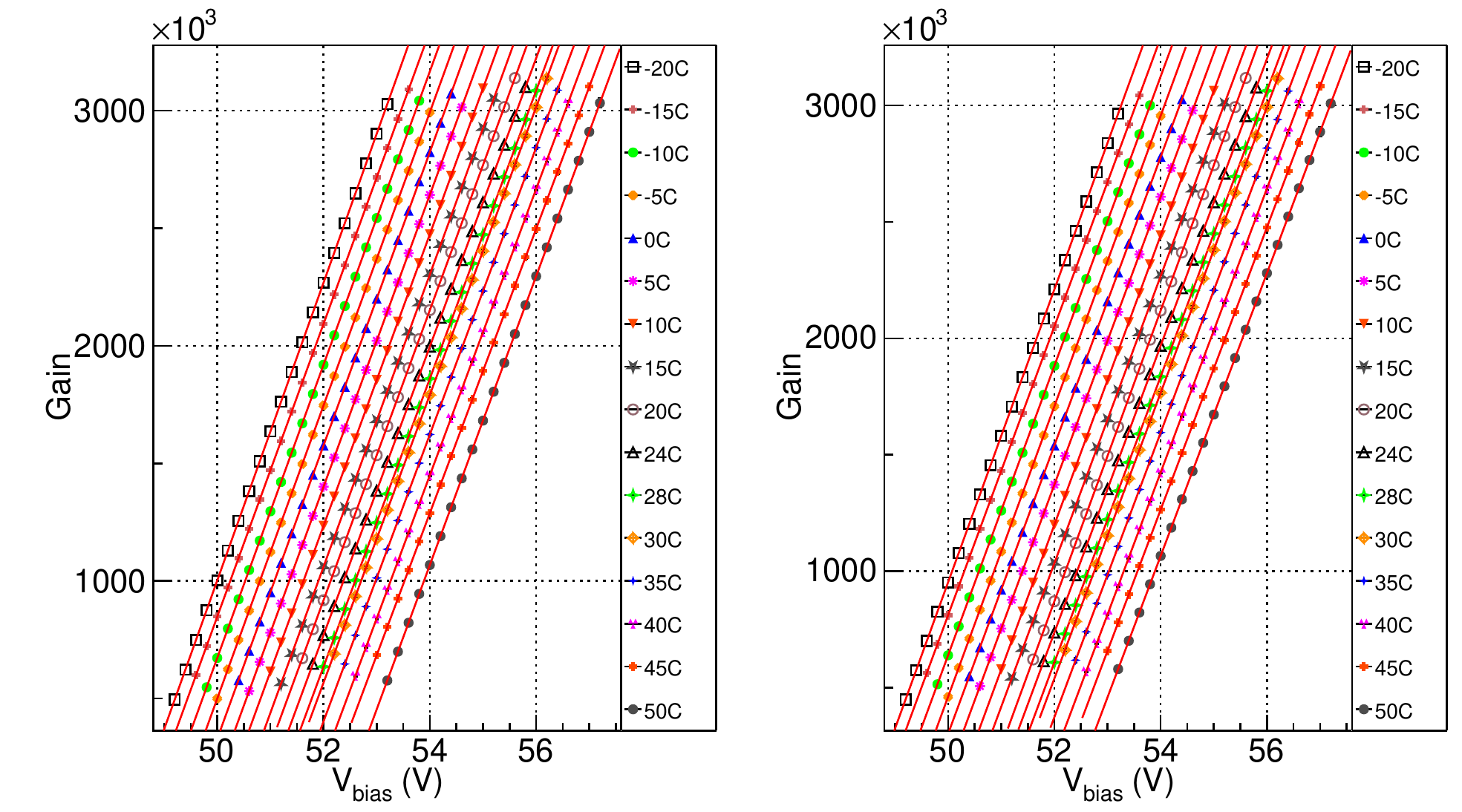}
\caption{The gain versus applied voltage for various temperatures for two SiPMs.}
\label{fig:4dgdV}
\end{figure} 

\begin{figure} [htbp]  
\centering
\includegraphics[width=0.8\textwidth]{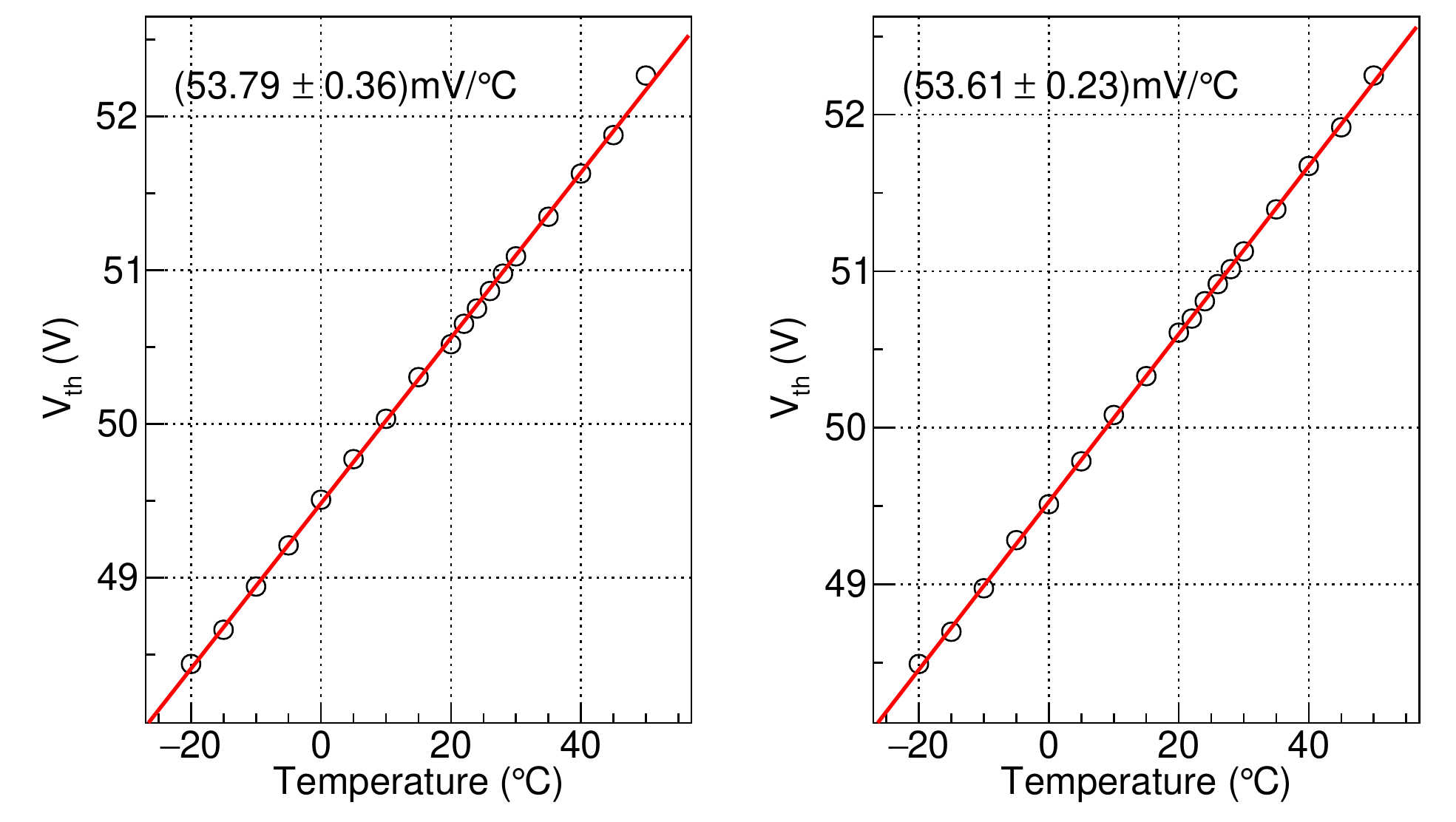}
\caption{Variation of $V_{th}$ with temperature for the same 2 SiPMs.}
\label{fig:4vbdt}
\end{figure}

The variations of absolute gain as a function of applied voltage are shown in figure~\ref{fig:4dgdV} for different temperatures. The gain of SiPM is varying linearly with increasing applied voltage and hence is fitted with a linear function for estimating the variation of gain with respect to the applied voltage (dG/dV) and threshold voltage ($V_{th}$) at each temperature. The variations of the observed $V_{th}$ as a function of temperature are shown in figure~\ref{fig:4vbdt} along with a linear fit to estimate the variation of $V_{th}$ with respect to temperature ($dV_{th}/dT$) for two of the SiPMs. The observed values of $dV_{th}/dT$ for tested SiPMs are consistent with the value given in specifications~\cite{sipmspecs}.

\begin{figure} [htbp]  
\centering
\includegraphics[width=0.8\textwidth]{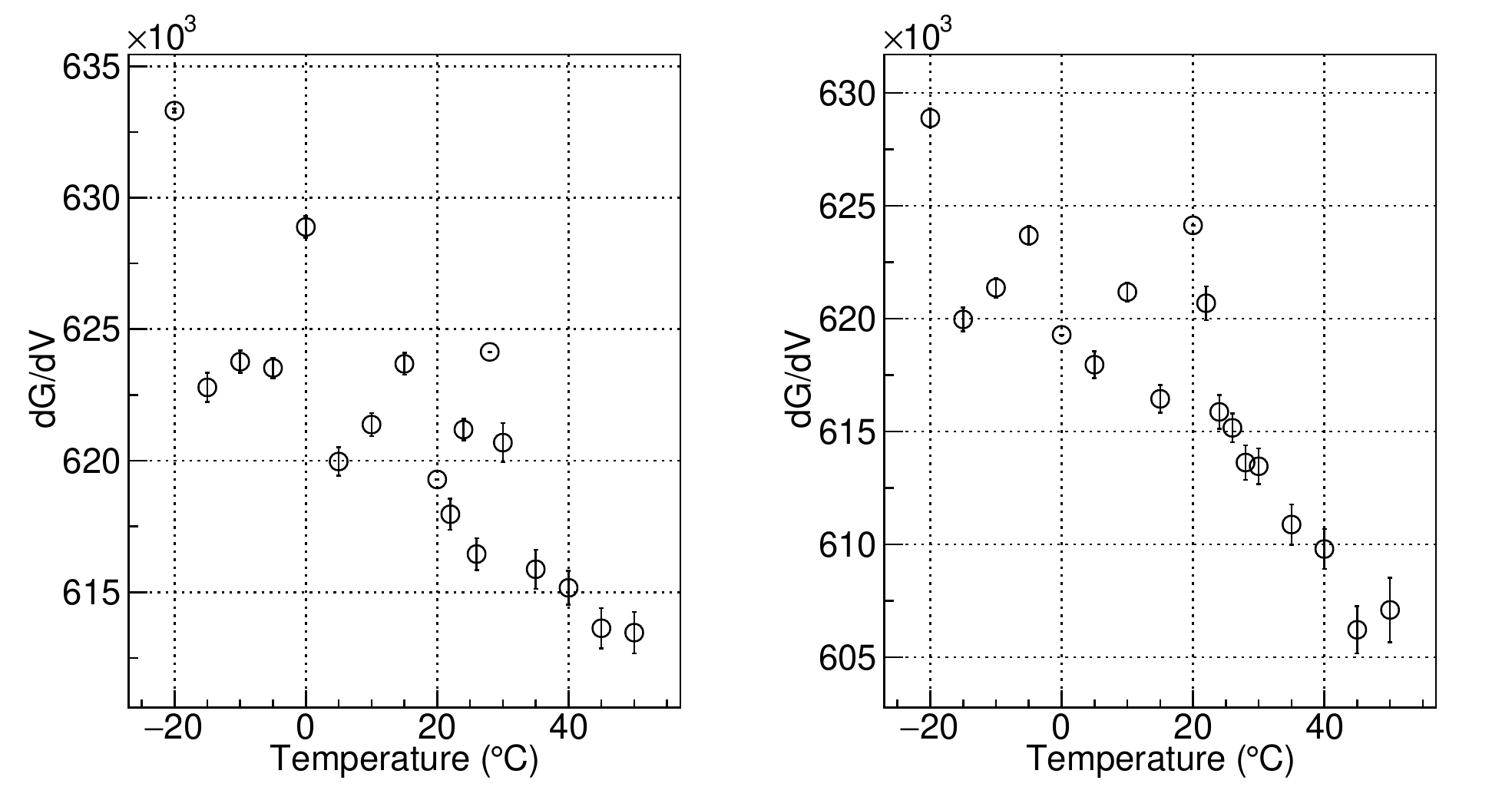}
\caption{$dG/dV$ versus temperature for two of the SiPMs.}
\label{fig:4dgdVvsT}
\end{figure}

The observed slopes ($dG/dV$) from  the linear fit in figure~\ref{fig:4dgdV} at different temperatures are shown in figure~\ref{fig:4dgdVvsT} for few SiPMs. These plots show the decrease of $dG/dV$ with temperature. The behaviour has been seen in literature~\cite{gainstablisation1,gainstablisation2}.

\begin{figure} [htbp]  
\centering
\includegraphics[width=\textwidth]{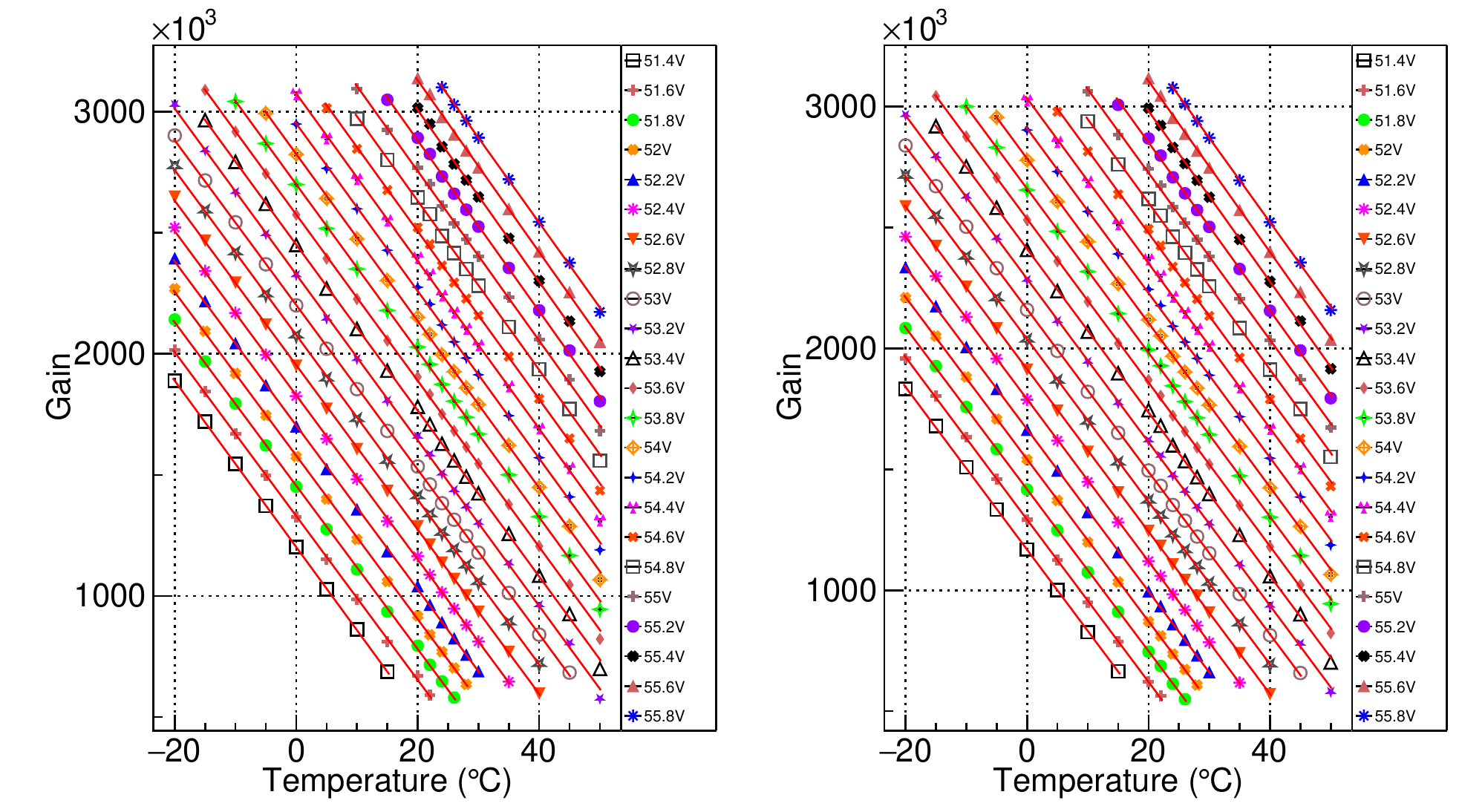}
\caption{The gain versus temperature for various applied voltages for the two SiPMs.}
\label{fig:4dgdt}
\end{figure}

\begin{figure} [htbp]  
\centering
\includegraphics[width=\textwidth]{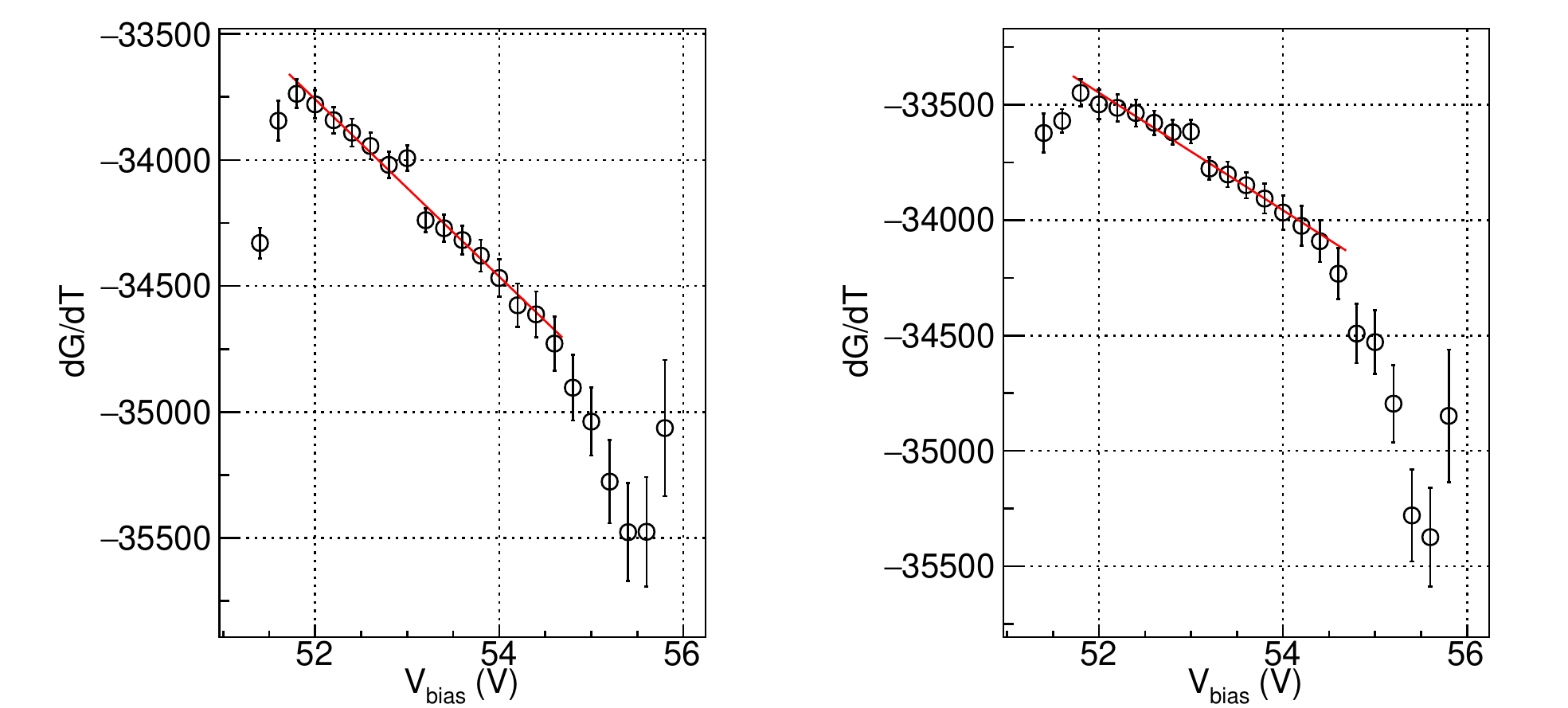}
\caption{$dG/dT$ versus applied voltage for two SiPMs.}
\label{fig:4dgdtvsV}
\end{figure}

The variations of absolute gain as a function of temperature are shown in figure~\ref{fig:4dgdt} for different applied voltages. As expected the gain is higher at lower temperatures, which is mainly due to the shift of $V_{ov}$. For each applied voltage the gain is fitted with a linear function and the observed slopes ($dG/dT$) for these two SiPMs are shown in figure~\ref{fig:4dgdtvsV}. The magnitude of $dG/dT$ varies with the applied voltage, which indicates a variation of $dG/dV$ for different temperatures.

\begin{figure} [htbp]  
\centering
\includegraphics[width=0.8\textwidth]{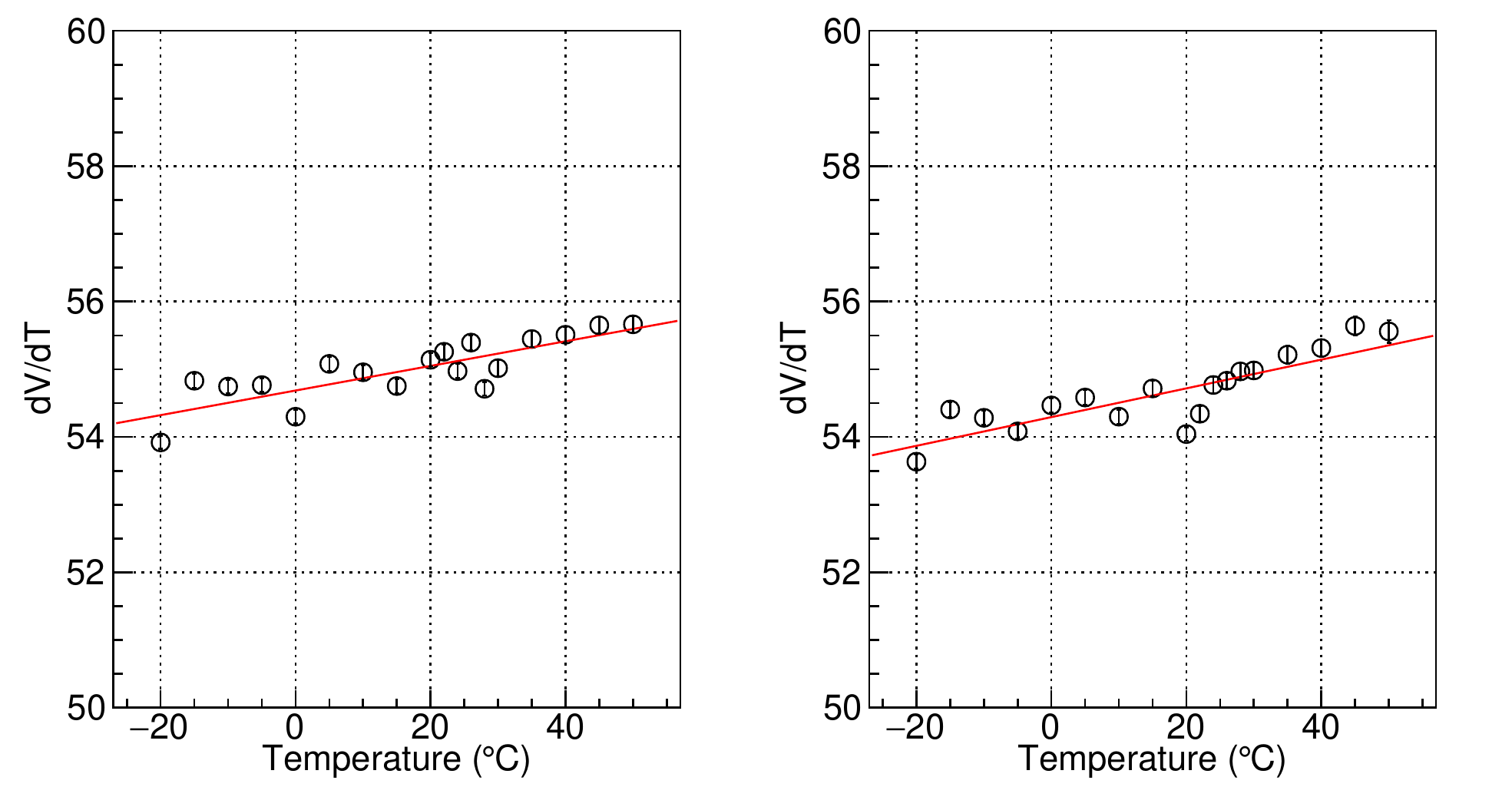}
\caption{dV/dT versus T for two of the SiPMs.}
\label{fig:4dvdTvsT}
\end{figure}

Using the slopes $dG/dV$ and $dG/dT$, $dV/dT$ is estimated. The average value of $dG/dT$ is evaluated using the values of $dG/dT$ in linear range from figure~\ref{fig:4dgdtvsV}. The ratio of average value of $dG/dT$ to $dG/dV$ at a given temperature is used to calculate $dV/dT$ at that particular temperature. The variations in $dV/dT$ are shown in figure \ref{fig:4dvdTvsT} as a function of temperature, where the value varies from (53.5 - 55.5)\,mV/$^\circ$\,C.

\section{The effect on efficiency and noise rate with the change in $V_{ov}$}
There is no extra control of temperature in the CMVD electronics, other than the default control through the room air conditioning system in the mini-ICAL room, where the variation of temperature is about 2$^{\circ}$\,C. The gain of the SiPM decreases with increasing temperature because $V_{th}$ shifts toward the higher side. The change in gain of SiPM with temperature will have an impact on the efficiency and the noise rate of the CMVD. To compensate for the gain change due to temperature change, the bias to the SiPM needs to be adjusted. As it can be observed in figure~\ref{fig:4dvdTvsT}, average value of $dV/dT$ is $\sim$55\,mV/$^\circ$\,C. This implies that the change in operating voltage would be $\sim$110\,mV. Do we need a feedback system in the SiPM power supply to compensate for the variation of temperature for the CMVD requirement, i.e. the efficiency of the veto wall should be more than 99.99\% with the range of temperature variation, and similarly the ratio of the fake to true muon rate should be less than $10^{-5}$ ? The randomly/noise triggered data and the cosmic muon triggered data have been collected at $V_{ov}$ of 2.5\,V $\pm$ 200\,mV at room temperature, whereas 2.5\,V is optimised value of $V_{ov}$ for CMVD operation~\cite{mamta1}. A total of 13 sets of data are collected to see the effect of gain change of SiPM (w.r.t. applied voltage change) on the CMVD performance.

\begin{figure} [htbp]  
\centering
\includegraphics[width=\textwidth]{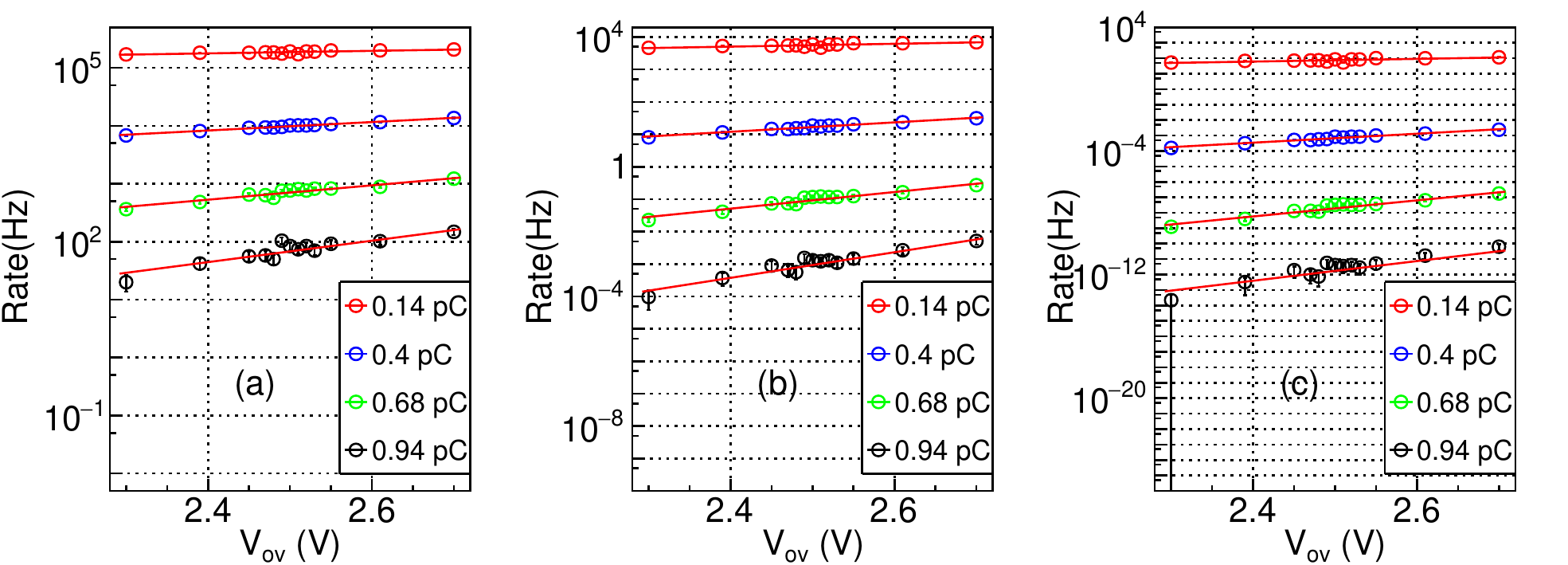}
\caption{The noise rate as a function of $V_{ov}$ and $q_{th}$ of (a) an individual SiPM, (b) a scintillator where at least two out of four SiPMs will have signal and (c) of CMVD for atleast two out of four layers satisfy trigger criteria when at least two out of four SiPMs will have signal in a scintillator.}
\label{fig:si_sc_la_noise}
\end{figure}

From the noise data, the individual noise rates of 4 SiPMs are calculated as a function of $V_{ov}$ and $q_{th}$ and is shown in figure~\ref{fig:si_sc_la_noise}(a) for one the SiPMs. Using the individual noise rates of 4 SiPMs, the noise rate of a scintillator is estimated when at least two out of four SiPMs have a signal and then these noise rates are propagated to estimate the noise rate of CMVD for any two out of four-layer trigger criterion as a function of $V_{ov}$ and $q_{th}$ and the rates are shown in figure~\ref{fig:si_sc_la_noise}(b) and figure~\ref{fig:si_sc_la_noise}(c) respectively. At the chosen $q_{th}$=0.68\,pC which is equivalent to 2.5\,p.e. for $V_{ov}$=2.5\,V, 
for CMVD operation, the noise rate of CMVD for the highest value of $V_{ov}$ (i.e. 2.7\,V) is $\sim$ $10^{-7}$\,Hz. The cosmic muon signal will be integrated within a 100\,ns during CMVD operation. The noise rate at $q_{th}$=0.68\,pC for a 100\,ns window will be $\sim$ $10^{-14}$\,Hz which is well below the tolerable noise rate for CMVD requirement~\cite{mamta1}.

\begin{figure} [htbp]
\centering
\includegraphics[height=5.25cm]{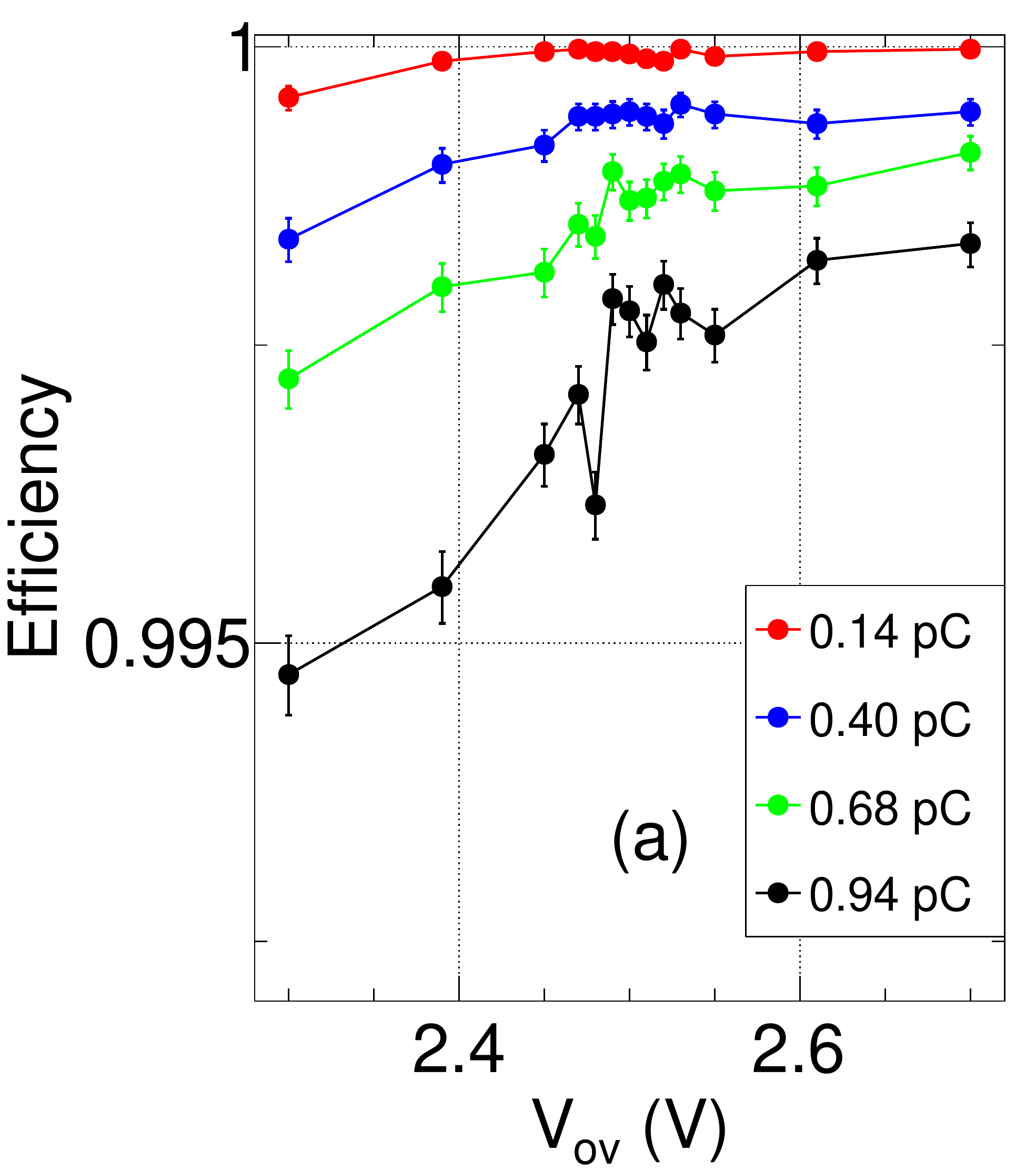}
\includegraphics[height=5.25cm]{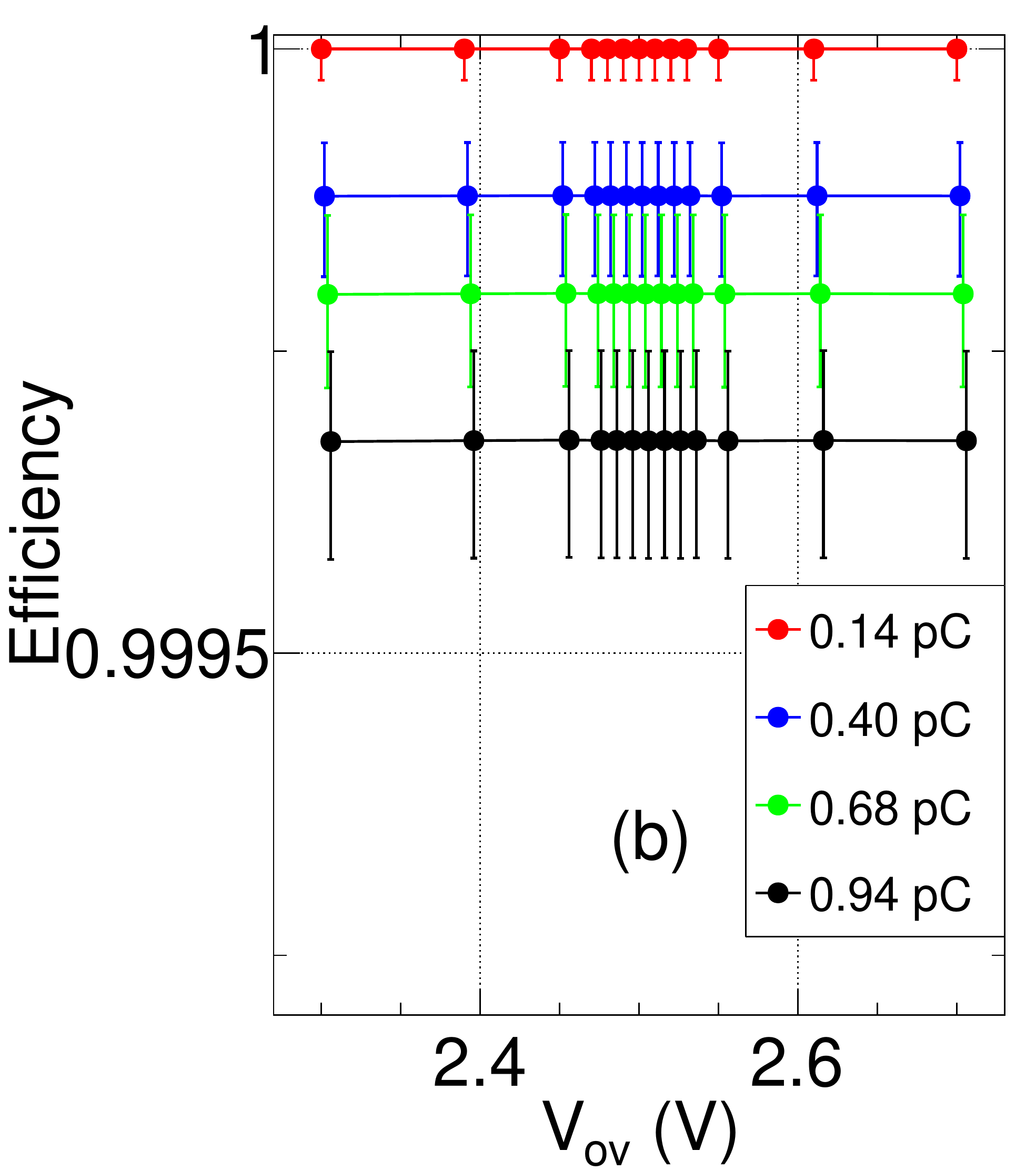}
\includegraphics[height=5.25cm]{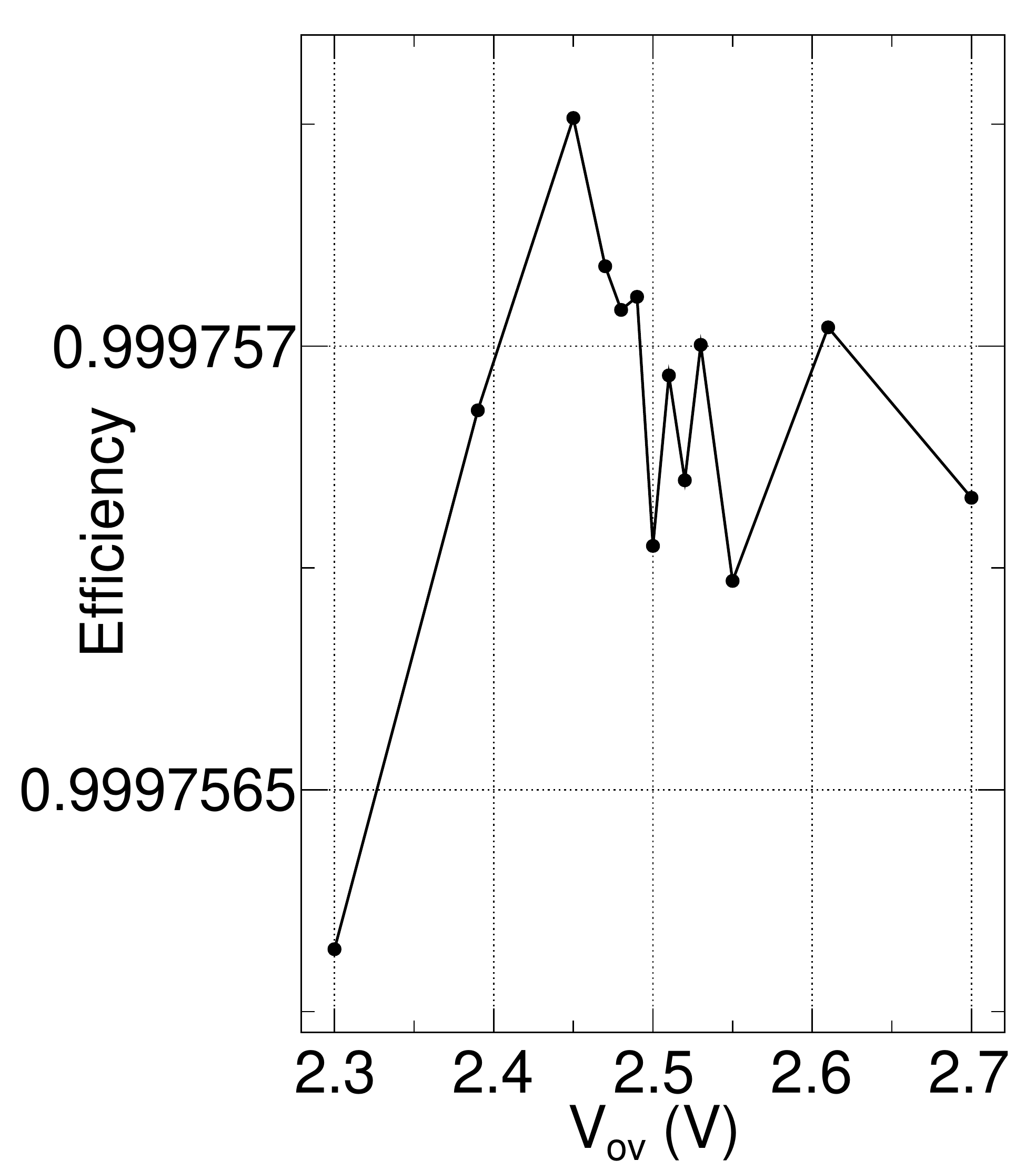}
\caption{The muon detection efficiency of (a) an individual SiPM as a function of $V_{ov}$ and $q_{th}$, (b) a scintillator as a function of $V_{ov}$ and $q_{th}$ when at least two out of four SiPMs will have signal and (c) a scintillator for $q_{th}$=0.68\,pC, the error on the data points in last plot is less than $10^{-4}$.}
\label{fig:si_sc_la_eff}
\end{figure}

For the efficiency measurement of a complete scintillator of length 4.7\,m, the scintillator is equipped with WLS fibre and SiPMs are mounted on both sides to collect the signal and measure the efficiency of the cosmic muon, which are triggered by three similar scintillator detectors. The applied voltage of the SiPMs is varied up to $\pm$200\,mV about the $V_{ov}$= 2.5\,V and the variation in the efficiency of SiPM and the
scintillator detector is checked and then extrapolated to the efficiency of the four-layer veto wall as described in~\cite{mamta1}.

The cosmic muon efficiency is measured for individual SiPMs as shown in figure~\ref{fig:si_sc_la_eff}\color{blue}(a)\color{black}\hspace{0.1cm} as a function of $V_{ov}$ and $q_{th}$. Also, the cosmic muon efficiency is measured when atleast two out of four SiPMs will have signal and it is shown in figure~\ref{fig:si_sc_la_eff}\color{blue}(b)\color{black}\hspace{0.1cm} as a function of $V_{ov}$ and $q_{th}$. Figure~\ref{fig:si_sc_la_eff}\color{blue}(c)\color{black}\hspace{0.1cm} shows the cosmic muon efficiency of a scintillator as a function of $V_{ov}$ at $q_{th}$=0.68\,pC, this plot shows the statistical fluctuations in efficiency for different $V_{ov}$ which can't be seen in figure~\ref{fig:si_sc_la_eff}\color{blue}(b)\color{black}\hspace{0.1cm} due to the efficiency scale. The scintillator efficiency at the lowest value of $V_{ov}$ i.e. 2.3\,V, is high enough to attain the efficiency requirements for the CMVD operation~\cite{mamta1}.

This study concludes that even if $V_{ov}$ is not compensated for $\pm$ $4^\circ$\,C variation at IICHEP lab, CMVD requirements will still be satisfied. Thus, neither it require a special temperature control on top of the air conditioning system nor a feedback option to change the operating voltage.

\section{Summary}
The main goal of this study is to test all SiPMs for the CMVD. The test results based on the gain and noise study show that all except one SiPM are acceptable for the CMVD. The variation in the threshold voltage of SiPM with temperature is studied and is found to be (53.5 - 54.5)\,mV/$^\circ$\,C. The CMVD will be operated in an air-conditioned room, where the maximum variation in temperature is 2$^\circ$\,C. Thus, a detailed study was done to find the variation in gain of SiPM as a function of temperature, which is found to be less than 2\%/$^\circ$\,C. This study also confirms that even with this variation of temperature, the performance of the CMVD detector will satisfy the design goal. The feedback electronic circuit will not be required to change the operating voltage to compensate for the variation of SiPM gain due to the variation in temperature.


\section{Acknowledgments}
We sincerely thank Prof. Varsha R Chitnis,  Mano Ranjan, S.R. Joshi, Darshana Gonji, Santosh Chavan, Vishal Asgolkar for their support and help during the work. We would also like to thank all other members of the INO collaboration for their valuable inputs.

\end{document}